\documentclass[%
 aip,
 amsmath,amssymb,
 reprint,%
]{revtex4-1}
\usepackage{graphicx}% Include figure files
\usepackage{dcolumn}% Align table columns on decimal point
\usepackage{bm}% bold math
\usepackage[utf8]{inputenc}
\usepackage[T1]{fontenc}
\usepackage{mathptmx}
\usepackage{etoolbox}
 
\usepackage{amsmath, amssymb,amsbsy}  % 用于数学公式
\usepackage{graphicx}         % 用于插入图片
\usepackage{dcolumn}          % 用于对齐表格中的小数点
\usepackage{bm}               % 用于粗体数学符号
\usepackage[utf8]{inputenc}   % 支持 UTF-8 编码
\usepackage[T1]{fontenc}      % 支持正确的字体渲染
\usepackage{mathptmx}         % 使用 Times 字体
\usepackage{etoolbox}         % 用于高级命令操作
\usepackage{subcaption}       % 用于插入图片
\usepackage{bm}
\usepackage{caption}

\begin{document}

\preprint{}

\title{Fluid-Structure Interaction and Scaling Laws for Deterministic Encapsulation of Hyperelastic Cells in Microfluidic Droplets}
% Force line breaks with \\
\author{Liu Andi}
%\altaffiliation[Also at ]{}%Lines break automatically or can be forced with \\
\author{Hu Guohui}
 \email{ghhu@staff.shu.edu.cn.}
\affiliation{Shanghai Institute of Applied Mathematics and Mechanics, School of Mechanics and Engineering Science, Shanghai Frontier Science Center
of Mechanoinformatics, Shanghai Key Laboratory of Mechanics in Energy Engineering, Shanghai University, Shanghai, China %\\This line break forced with \textbackslash\textbackslash
}%

\begin{abstract}
  
The precise encapsulation of deformable particles in multiphase flows involves complex transient Fluid-Structure Interactions (FSI) and topological interfacial changes. In the context of single-cell analysis, a numerical framework that couples the Cahn-Hilliard phase-field model with the Arbitrary Lagrangian-Eulerian (ALE) method is employed to investigate the dynamics of deformable cell encapsulation in flow-focusing microchannels. By resolving the coupling between the hyperelastic cell, carrier fluid, and evolving interface, we propose a unified dimensionless scaling law to predict the operational spatial window for the deterministic encapsulation quantitatively. Furthermore, the physical presence of cells modulates the droplet generation flow regime via a "geometric blockage effect", shifting the transition boundary from the squeezing to the dripping regime toward lower flow-rate ratios. The droplet generation period demonstrates a non-monotonic dependence on the cell blockage ratio $\Gamma$, which induces a competitive mechanism between shear enhancement and hydraulic resistance penalty, and consequently leads to an optimal hydrodynamic balance at $\Gamma \approx 0.32$. Finally, we find that while droplet periodicity is robust to variations in cell stiffness, the transient stress field within the cell is highly sensitive, particularly during the capillary pinch-off singularity. This work clarifies the fundamental interaction between hyperelastic cells and multiphase flows, and provides a quantitative framework for optimizing damage-free cell encapsulation systems.

\end{abstract}

%\keywords{Flow focusing; three-phase flow; fluid-solid coupling; droplet generation; cell encapsulation.}

\maketitle

\section{\label{sec:level1}Introduction}

The precise manipulation of deformable soft particles (e.g., hyperelastic cells) in multiphase microflows is a fundamental problem in fluid mechanics with significant engineering applications in high-throughput single-cell analysis. As a breakthrough tool in life science research, single-cell analysis has shown great promise in biomedical fields such as cancer research \cite{Carneiro2025, Jiang2024}, neuroscience \cite{Tasic2018}, and stem cell biology \cite{Fan2019} by overcoming the limitations of traditional population analysis \cite{loo2007}.  
Microfluidic platforms are emerging as a crucial technology in single-cell analysis, owing to their advantages of low reagent consumption, minimized cross-contamination between cells, high detection sensitivity, and large-scale integration of subsystems \cite{Jiang2023}

As a critical step in the single-cell analysis procedure, encapsulating deformable cells into droplets is enabled by microfluidics-based high-throughput technologies, such as flow focusing or co-flow structures \cite{Utech2015, Rajeswari2017, Zhang2018, Ng2016, Macosko2015}. However, stochastic cell encapsulation in flow focusing inherently suffers from a high occurrence of empty droplets and multi-cell occupancy, i.e., less than 35\% of droplets were found to contain exactly one particle\cite{lagus2013}.{\it } Multi-dimensional optimization strategies have been extensively utilized to improve the encapsulation efficiency to $70\%\sim 80\%$, such as inertial sorting\cite{Edd2008}, Dean flow\cite{Kemna2012}, or cell-triggered Rayleigh–Plateau instability\cite{Chabert2008}. Nevertheless, deterministic “on-demand” encapsulation, i.e., ensuring exactly one cell per droplet, remains a formidable challenge. 
Therefore, understanding the non-linear dynamic behavior of these deformable cells in multiphase flows is critical for optimizing the hydrodynamic performance of such systems.

The droplet generation in two-phase flows within microchannels governs the dynamics of the encapsulation process. Guillot and Colin\cite{Guillot2005} experimentally observed three different flow regimes in a microfluidic T-junction. Depending on the distinct flow parameters (e.g., Capillary number $Ca$ and flow rate ratio $Q$), typical flow patterns such as squeezing (Droplet at Cross-Junction, DCJ), dripping (Droplet in Co-flowing, DC), and jetting (Parallel Flow, PF) have been identified\cite{Tan2008, Liu2011, Garstecki2006, Menech2008}. 
Using a Boltzmann multiphase lattice model, Liu et al.\cite{Liu2011} observed the above three flow regimes through three-dimensional numerical simulation, and systematically studied the effects of the flow rate ratio, capillary number, and channel geometry in the DCJ region dominated by squeezing pressure. 
Based on the two types of fragmentation mechanisms, namely extrusion \cite{Garstecki2006} and shear-driven \cite{thorsen2001}, Menech et al.\cite{Menech2008} clarified the underlying physics of the three droplet regimes through numerical studies on T-shaped microchannel flows. 
Fu et al.\cite{Fu2012} experimentally investigated the droplet formation and breakup dynamics under dripping or jetting conditions in a flow-focusing microchannel. Castro-Hernández et al.\cite{Castro2016} numerically and experimentally studied the influence of geometric parameters on droplet formation in a three-dimensional flow-focusing device. Sontti et al.\cite{Sontti2023} systematically investigated the effects of contraction width, length, continuous phase flow rate, and interfacial tension on droplet size, velocity, shape, and volume using a three-dimensional unsteady Computational Fluid Dynamics (CFD) model for droplet formation.

To understand the dynamics of droplet cell encapsulation, numerical simulation has proven effective in solving Fluid-Structure Interaction (FSI) problems with moving boundaries. Yang et al.\cite{Yang2013} employed the lattice Boltzmann method combined with the immersed boundary method to simulate the cell encapsulation as well as the cell deformation inside the droplet, in which cell deformation was described using a hybrid membrane model that accounts for in-plane tension and bending stiffness via a spring network.
Yaghoobi et al.\cite{Yaghoobi2020} employed the finite element method in COMSOL Multiphysics software to simulate the motion of rigid particles in a flow-focusing channel and revealed that encapsulation is highly sensitive to inter-particle spacing.
Fatehifar et al.\cite{Fatehifar2023} utilized open-source software to explore the impact of particle pre-sorting on rigid particle encapsulation in a high-throughput flow-focusing microchannel under different flow modes. 
Jannati et al. \cite{Jannati2025} employed the lattice Boltzmann method coupled with the immersed boundary method, and modeled the cell as a deformable membrane governed by Hookean spring networks and Helfrich bending potentials. They focused on the cell-induced mixing dynamics in the encapsulating droplet across varying fluid and material parameters.

Although the numerical studies available contributed to understanding the basic coupling between flow fields and particle motion, they typically simplified biological cells as rigid particles or membrane-based spring networks. They have limitations in capturing the complex material nonlinearity and large deformation dynamics governed by hyperelastic constitutive laws.
Consequently, the fundamental mechanisms governing common experimental failures, such as channel blockage induced by geometric constraints or the "premature escape" of highly deformable cells, remain mechanistically obscure. More importantly, the transient stress evolution within the cell during the violent topological transition of the droplet necking stage has rarely been quantified. This lack of detailed stress analysis hinders the assessment of mechanical damage, which is critical for ensuring the biosafety and functional viability of encapsulated cells. Furthermore, a quantitative theoretical framework, specifically dimensionless scaling laws that predict successful encapsulation with multiphase flow parameters, is critically needed to provide guidelines for the robust engineering design of microfluidic devices.

In the context of single-cell analysis, this study examines the complex fluid-structure interaction of hyperelastic cells using a fully coupled numerical framework that integrates the Arbitrary Lagrangian-Eulerian (ALE) formulation with the phase-field method. In contrast to the membrane model\cite{Yang2013, Jannati2025}, our study describes the cell as a continuum hyperelastic solid, which is essential for accurately predicting the internal stress transmission and strain energy storage within the cell body for assessing cell safety. Through numerical simulation, we aim to understand the transient hydrodynamic mechanisms, particularly the retraction vortex induced by the interface breakup; to quantify the non-linear stress evolution within the cell for assessing mechanical risks; and to derive dimensionless scaling laws that predict the critical spatial thresholds for deterministic encapsulation. These findings provide a solid theoretical basis for the design and optimization of microfluidic systems.

\section{\label{sec:level2}Governing Equations}

For the oil-water-solid three-phase coupled model in simulating cell encapsulation in microchannels, this study utilizes the phase field model \cite{two-phase1,two-phase2,two-phase3,two-phase4,two-phase5} to describe the formation of droplets in microfluidic intersections, and employs the ALE fluid-solid coupling \cite{ALE1, ALE2, ALE3, ALE4, ALE5, ALE6, ALE7} method to investigate the interaction between droplets and cells.

\subsection{\label{sec:level2}Phase-field Equation}

The phase field method is a widely used numerical approach for simulating the evolution of interfaces in multiphase flows. It assumes that the interface between the phases is diffusive, and the phase field variables change continuously at the interface while remaining uniform throughout each fluid phase. The evolution of the two-phase interface in the phase field model is typically described by the Cahn-Hilliard equation, which is based on the principle of minimizing free energy \cite{Cahn1958}. It represents the phase separation process by considering the competition between interfacial energy and bulk energy. 
  
The total free energy $F(\phi)$ of the mixed region of the two-phase interface \cite{Cahn1959} is expressed as follows:
  \begin{eqnarray}
    F(\phi) = \int_{\Omega} [f_1(\phi) + f_2(\phi)] dV \label{eq:one}.
  \end{eqnarray}
Here, $\Omega$ represents the volume of the entire system, and $\phi$ is the phase field parameter. The first term in the right-hand side integral represents the interfacial energy, which is used to limit the spatial rate of change of the order parameter (to prevent the interface from spreading infinitely). Usually, the gradient squared term is adopted. The formula for the interfacial energy is:
  \begin{equation}
  f_1(\phi) = \frac{\varepsilon}{2} (\nabla \phi)^2,
  \end{equation}
in which $\varepsilon$ stands for the interface thickness parameter. The second term is the bulk energy, and its most common form is the Ginzburg-Landau double-well potential, which is used to describe the stability of the two phases \cite{two-phase1}. The formula for the double-well potential function is:
  \begin{equation}
  f_{2}(\phi) = W(\phi^{2} - 1)^{2},
  \end{equation}
where $W$ is the height of the double-barrier potential well and serves as a measure of the free energy difference of the control volume.
  
The chemical potential $G$ is defined as the functional derivative of the total free energy:
  \begin{equation}
   G = \frac{\delta F(\phi)}{\delta \phi} = -\epsilon \nabla^{2} \phi + 4W\phi(\phi^{2} - 1).
   \end{equation}
Based on the above formula and the law of conservation of mass, the Cahn-Hilliard equation can be derived,
   \begin{equation}
   \frac{\partial \phi}{\partial t} + \nabla \cdot (\phi \mathbf{v}^{f}) = M\nabla^{2}G,
   \end{equation}
in which $t$ represents time, $M$ is the mobility, and $\mathbf{v}^{f}$  is the fluid velocity.
 
\subsection{\label{sec:level2}Incompressible Navier-Stokes Equations}

The fluid surrounding the cells is considered to be an incompressible two-phase Newtonian fluid. The laws of momentum conservation and mass conservation constrain the fluid region. In the present study, the fluid interface is simulated using the Navier-Stokes equations with surface tension. The governing equations on the moving grid can be written as,
  \begin{equation}
  \left\{\begin{array}{l}\rho(\phi)\left(\displaystyle\frac{\partial \mathbf{v}^{f}}{\partial t}+(\mathbf{v}^{f}-\mathbf{v}^{g})\cdot \nabla \mathbf{v}^{f}\right)=\nabla \cdot \boldsymbol{\sigma}^{f} + \mathbf{F}_{\gamma} \\ \nabla \cdot \mathbf{v}^{f}=0\end{array},\right.
  \end{equation}
in which,
  \begin{equation}
  \rho(\phi) = \rho_{c}V_{c}^{f} + \rho_{d}V_{d}^{f},
  \end{equation}
  \begin{equation}
   \mu(\phi) = \mu_{c}V_{c}^{f} + \mu_{d}V_{d}^{f},
  \end{equation}
  \begin{equation}
   V_{c}^{f} = \frac{1-\phi}{2},\ V_{d}^{f} = \frac{1+\phi}{2},\ V_{c}^{f} + V_{d}^{f} = 1.
  \end{equation}
In the equations, $\phi=-1$ represents the oil phase, and $\phi=1$ represents the water phase. $V_{c}^{f}$ and $V_{d}^{f}$ represent the volume fractions of the oil phase and the water phase, respectively. $\rho_{c}$ and $\rho_{d}$ are the densities of the oil phase and the water phase. $\mu_{c}$ and $ \mu_{d}$ denote the dynamic viscosities of the oil phase and the water phase, respectively. $\mathbf{v}^{g}$ is the grid velocity, and ${\sigma}^{f}$ is the total stress tensor of the fluid,
  \begin{equation}
   \boldsymbol{\sigma}^{f} = -p\mathbf{I} + \mu(\phi) \left( \nabla \mathbf{v}^{f} + (\nabla \mathbf{v}^{f})^{T} \right).
  \end{equation}
Here, $p$ represents the static pressure of the fluid, which reflects the isotropic pressure of the liquid, and $\mathbf{I}$ is a unit tensor. 
%$\mathbf{F}_{\sigma}$ is the surface tension of oil-water interface. 
The Continuous Surface Force (CSF) model, proposed by Brackbill et al. \cite{Brackbill1992} and introduced into the phase field method by Liu et al. \cite{Liu2017}, is employed to describe the interfacial dynamics.
This not only simplifies the calculation of the interface curvature but also introduces the tangential surface tension and its variations,
  \begin{equation}
  \mathbf{F}_{\gamma} = \frac{3\sqrt{2}}{4} \epsilon \left[ \frac{\gamma}{\epsilon^{2}} \mu(\phi)\nabla \phi + |\nabla \phi|^{2} \nabla \gamma - (\nabla \gamma \cdot \nabla ) \nabla \phi \right],
  \end{equation}
in which the first term within the right parenthesis represents the normal component of surface tension, and the last two terms on the right side stand for the tangential components, namely the Marangoni Stress. When the surface tension is uniformly distributed along the interface or the surface tension coefficient is constant, the tangential surface tension becomes zero.

\subsection{\label{sec:level2}Cell Constitutive Equation}

Cells are described using the hyperelastic Mooney-Rivlin model in the present study. This model has been applied in numerous research fields related to biological modeling \cite{Mallinson2004,soleimani2016,xue2024, Wang2015, Gong2024}, such as cell squeezing techniques \cite{Xie2021}, cell cavitation deformation \cite{Wang2015}, and prediction of the cell elastic modulus \cite{Gong2024}.

The momentum conservation equation for the solid is given by:
  \begin{equation}
  \rho^{s} \frac{\partial^{2} \mathbf{u}^{s}}{\partial t^{2}} - \nabla \cdot \boldsymbol{\sigma}^{s} = 0,
  \end{equation}
in which $\rho^{s}$ represents the solid density, $\mathbf{u}^{s}$ represents the solid displacement, and ${\sigma}^{s}$ represents the Cauchy stress tensor, which is defined by its constitutive relationship.

The strain energy density function of the M-R model is as follows:
  \begin{equation}\label{MR}
  W = C_{1}(J_1 - 3) + C_{2}(J_2 - 3) + \frac{\kappa_{c}}{2} (J_3 - 1)^2,
  \end{equation}
where $C_{1}$ and $C_{2}$ are material constants, representing parameters that characterize the elastic modulus of the cells. $\kappa_{c}$ is the bulk modulus. $J_1$, $J_2$, and $J_3$ are invariants related to the deformation gradient, which are associated with the strain invariants $I_{1}$, $I_{2}$, and $I_{3}$, respectively.
  \begin{equation}
  J_{1} = I_{1}I_{3}^{-\frac{1}{3}},\ 
  J_{2} = I_{2}I_{3}^{-\frac{2}{3}},\ 
  J_{3} = I_{3}^{\frac{1}{2}},
  \end{equation}
  \begin{equation}
  I_{1} = \mathcal{C}_{kk},\ 
  I_{2} = \frac{1}{2}(I_{1}^{2} - \mathcal{C}_{ij}\mathcal{C}_{ji}),\
  I_{3} = \mathrm{det}({\pmb{\mathcal{C}}}).
  \end{equation}
The Cauchy-Green deformation tensor $\pmb{\mathcal{C}}$ is defined as $\pmb{\mathcal{C}} = \mathbf{F}^T \mathbf{F}$. $\mathbf{F}$ is the deformation gradient tensor, which describes the deformation mapping from the reference configuration (initial position $x_i$) to the current configuration (current position $X_j$), 
%and is the key tensor that connects the "initial - current" configurations.
  \begin{equation}
  \mathbf{F} = \begin{bmatrix} F_{xx} & F_{xy} \\ F_{yx} & F_{yy} \end{bmatrix},\  \ 
  F_{ij} = \frac{\partial x_{i}}{\partial X_{j}}.
  \end{equation}
  The Cauchy stress tensor can be calculated using,
  \begin{equation}
  \mathbf{\sigma}^{s} = J^{-1} \mathbf{F} \mathbf{S} \mathbf{F}^{T},
  \end{equation}
where $J = \mathrm{det}(\mathbf{F})$ and $\mathbf{S}$ is the second Piola-Kirchhoff stress tensor, which is given by:
  \begin{equation}
  \mathbf{S} = \mathbf{F}^{-1} \frac{\partial W}{\partial F_{ij}}.
  \end{equation}

\subsection{\label{sec:level2}Arbitrary Lagrange-Euler Method}

The ALE method combines the Lagrangian and Eulerian equations of fluid motion and has been widely utilized for simulating fluid-solid coupling problems. In this method, the solid grid follows the Lagrangian equation and can change over time, while the fluid grid uses the Eulerian equation and remains unchanged. The ALE method can handle the relative motion between the fluid and the solid, such as the movement and deformation of the solid, and the deformation and update of the grid can better adapt to the interaction between the fluid and the solid \cite{ALE2}.

The velocity of the moving grid ${\bf v}^{g}$ is defined by the grid displacement control equation:
  \begin{equation}
  \frac{\partial {\bf x}_{g}}{\partial t} = {\bf v}^{g},
  \end{equation}
  \begin{equation}
  \nabla \cdot {\bf v}^{g} = 0.
  \end{equation}
Among them, ${\bf x}_{g}$ represents the coordinate of the grid node.

\subsection{\label{sec:level2}Fluid-Solid Coupling Interface Conditions}
  
At the fluid-solid interface, the no-slip condition is imposed, that is, the fluid velocity is equal to the velocity of the solid interface:
  \begin{equation}
  \mathbf{v}^{f} = \frac{\partial \mathbf{u}^{s}}{\partial t}.
  \end{equation}
The force balance equation is satisfied, that is, the normal stress is continuous:
  \begin{equation}
  \boldsymbol{\sigma}^{f} \cdot \mathbf{n} = \boldsymbol{\sigma}^{s} \cdot \mathbf{n},
  \end{equation}
in which $\boldsymbol{\sigma}^{f}$ represents the fluid stress tensor, $\boldsymbol{\sigma}^{s}$  represents the solid stress tensor, and $\mathbf{n}$ represents the normal vector of the interface.
  
\section{\label{sec:level2}Numerical Implementation}

\subsection{\label{sec:level2}Discretization Method}

Numerical simulations are conducted for the two-dimensional governing equations in the present study. Although 3D simulations provide a more realistic representation, the strong coupling in three-phase flows introduces significant numerical stiffness to the governing equations, which strictly limits the time step. Furthermore, accurate modeling requires capturing multi-scale features, ranging from the channel width down to the thin liquid layer between the cell and the wall. Achieving such high resolution in 3D requires tremendous computational costs, making systematic parameter studies infeasible. To address this, we prioritize grid resolution and physical accuracy over geometric dimensionality. Validated by the previous 3D numerical results\cite{Liu2011} (see Fig.\ref{fig:2}), this study employs a high-resolution 2D model. This approach enables a detailed analysis of transient mechanisms (e.g., vortex structures and cell deformation) and a broad exploration of the parameter space (such as capillary number, blockage ratio, and stiffness) to derive reliable scaling laws.

The incompressible fluid and hyperelastic constitutive solid in the model are solved using the finite element method (FEM) framework \cite{Khoei2007,lopes2021,liu2022}, with spatial discretization achieved through the Galerkin weak form, using a second-order upwind scheme for the convection term to enhance stability in high flow velocity regions. The diffusion term is discretized using the central difference scheme to ensure the accuracy of the numerical calculation. First-order implicit backward difference and adaptive time stepping are adopted for time discretization of the stiff equations.

Full pressure-velocity coupling is based on solving a large system of coupled equations, simultaneously solving velocity and pressure unknowns such that velocity and pressure satisfy the continuity and momentum conservation equations at each time step, and thus ensures the accuracy of strong coupling.

Fluid-solid coupling is achieved through a two-way strong coupling for FEM-FEM discretization (incompressible fluid flow and hyperelastic solid constitutive model). 
Interfacial forces and displacement transmission satisfy the force balance condition through the penalty function method, and time coupling adopts a sub-iteration strategy (within a macroscopic time step, the "fluid-solid" alternating solution is carried out multiple times) to ensure the stability of the transient coupling process.

\subsection{\label{sec:level2}Computational Geometry}
 
\begin{figure*}[ht]
\captionsetup{justification=raggedright,singlelinecheck=false}
    \centering  % 只让图片居中
    \includegraphics[width=0.7\textwidth]{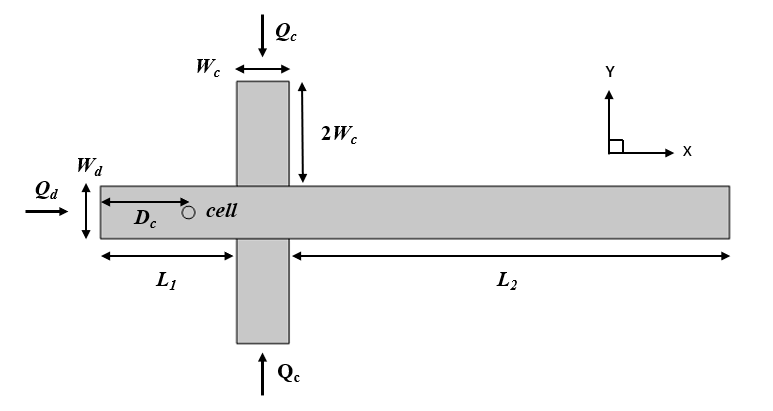}
    \caption{Schematic diagram of the cross-linked microchannel containing cells. $W_d$ and $W_c$ represent the widths of the main channel and the side channel, respectively, and the length of the side channel is $2W_c$. $L_1$ and $L_2$ represent the lengths of the main channel on the dispersed phase inlet side and the main channel on the continuous phase outlet side. $Q_c$ and $Q_d$ are the inlet volumetric flow rates of the continuous phase and the dispersed phase. The distance from the center of the cell to the inlet in the $X$ direction is $D_c$.}
    \label{fig:1} 
  \end{figure*}

\begin{table*}[ht]
\centering
\caption{Physical and geometric parameters in the present model}
\label{tab:1}
% 将表格宽度设置为页面文本宽度
\begin{tabular*}{\textwidth}{@{\extracolsep{\fill}} c c c c @{}}
\hline
\textbf{Parameter} & \textbf{Symbol/Formula} & \textbf{Values} & \textbf{Unit} \\
\hline
  Dispersed phase density & $\rho_{d}$ & 1000 \cite{Raj2010} & kg/m$^3$ \\
  Cell density & $\rho_{cell}$ & 1000\cite{Xie2021} & kg/m$^{3}$ \\
  Continuous phase density & $\rho_{c}$ & 1870\cite{Pu2024} & kg/m$^3$ \\
  Dynamic viscosity of the dispersed phase & $\mu_{d}$ & 0.001\cite{Raj2010} & Pa$\cdot$s\\
  Interfacial tension coefficient & $\gamma$ & 0.06\cite{Yaghoobi2020} & N/m\\
  Ratio of inlet widths & $ \Lambda=W_{d} / W_{c} $ & 1$\sim$2 & /\\ 
  Ratio of inlet volumetric flow rate & $ Q = Q_d / Q_c $ & 0.2$\sim$4.6 & /\\
  Viscosity ratio & $ \lambda = \mu_{d} / \mu_{c} $ & 0.1$\sim$0.3 & /\\
  Capillary number & $ Ca = \mu_{c} Q_{c} / W_{c}^{2} \gamma $ & 0.0006$\sim$0.008 & /\\
  Reynolds number & $ Re = \rho_{c} Q_{c} / \mu_{c} W_{c} $ & 0.1$\sim$0.4 & /\\
  Cell blockage ratio & $ \Gamma=2 R / W_{d} $ & 0$\sim$1 & /\\
  Cell material parameters & $ C_{1} $ & 300$\sim$1000 \cite{Mathur2000}& Pa\\
  Cell material parameters & $ C_{2} $ & 200\cite{Mathur2000} & Pa\\
  Length of the main passage on the entrance side & $ L_{1} $ & 100$\sim$500 & $\mu$m\\
  Length of the main passage on the exit side & $ L_{2} $ & 500$\sim$1000 & $\mu$m\\
  Distance between the cell and the inlet of the main channel & $ D_{c} $ & 10$\sim$500 & $\mu$m\\
\hline
\end{tabular*}
\end{table*}
 
The two-dimensional geometry of the cross-linked microchannel is illustrated in Fig. \ref{fig:1}. The microchannel consists of a main channel with a width of $W_d$ and two lateral channels with a width of $W_c$. The water-based dispersed phase is introduced at the entrance of the main channel, and the continuous phase oil is injected into the lateral channels. We assume that the fluid is single-phase at the inlet and outlet, and the microchannel outlet is a static pressure outlet.
To consider the interaction between the fluid and the solid surface, the contact angle of the hydrophilic cell wall is $60^{\circ}$, and the contact angle of the non-slip wall surface of the channel is $150^{\circ}$.

The analysis of fluid dynamics in flow-focusing junction microchannels can be described using five independent dimensionless numbers. The capillary number $Ca = \mu_{c} Q_{c} / W_{c}^{2} \gamma$ describes the relative dominance of viscous force and interfacial tension and is an important parameter for droplet formation. The Reynolds number $Re = \rho_{c} Q_{c} / \mu_{c} W_{c}$ is the most commonly used dimensionless number for describing microfluidics, and it is a measure of the ratio of inertial force to viscous force. During the droplet formation process, the continuous phase and the dispersed phase are continuously injected at different volumetric flow rates, and the inlet volumetric flow rate ratio $Q = Q_d / Q_c$ and the dynamic viscosity ratio $\lambda = \mu_{d} / \mu_{c}$ are two important dimensionless numbers characterizing droplet generation. The ratio of the inlet widths of the two-phase microchannel is determined by $\Lambda=W_{d} / W_{c}$. The geometric and physical parameters are listed in Table I, including inlet channel widths ($W_d$ and $W_c$), inlet volumetric flow rates ($Q_d$ and $Q_c$), dynamic viscosities ($\mu_{d}$ and $\mu_{c}$), interfacial tension coefficient ($\gamma$), and densities ($\rho_{d}$ and $\rho_{c}$), where the subscripts "$d$" and "$c$"  respectively represent the dispersed phase and the continuous phase. 
  
\begin{figure*}[ht]
 
    \centering  % 只让图片居中\centering 
    % 这是一个包含两张子图的大"图2"
  \begin{subfigure}{0.33\textwidth}
  \centering 
  \includegraphics[width=\textwidth]{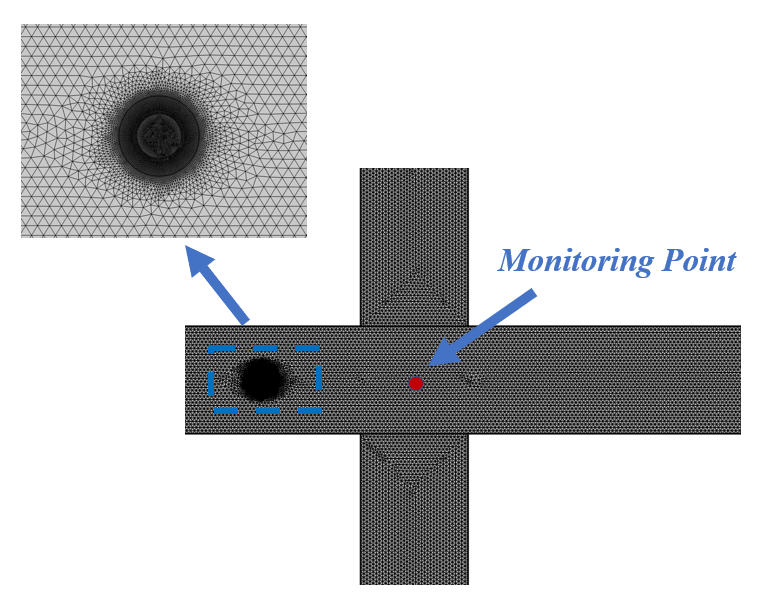}
  \caption{}
  \label{fig:2a}
  \end{subfigure}
  \hfill
  \begin{subfigure}{0.33\textwidth}
  \centering 
  \includegraphics[width=\textwidth]{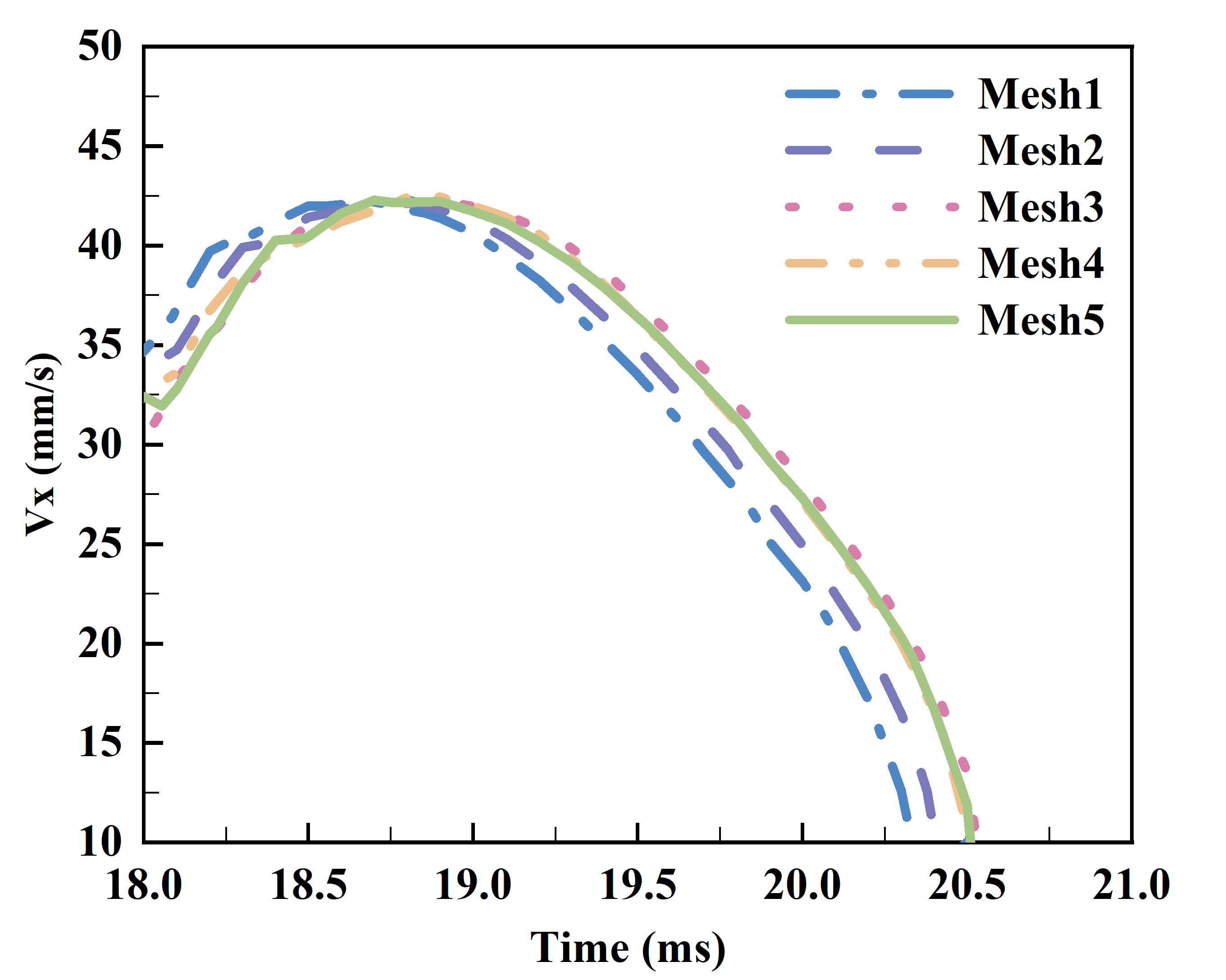}
  \caption{}
  \label{fig:2b}
  \end{subfigure}
  \hfill
  \begin{subfigure}{0.33\textwidth}
  \centering 
  \includegraphics[width=\textwidth]{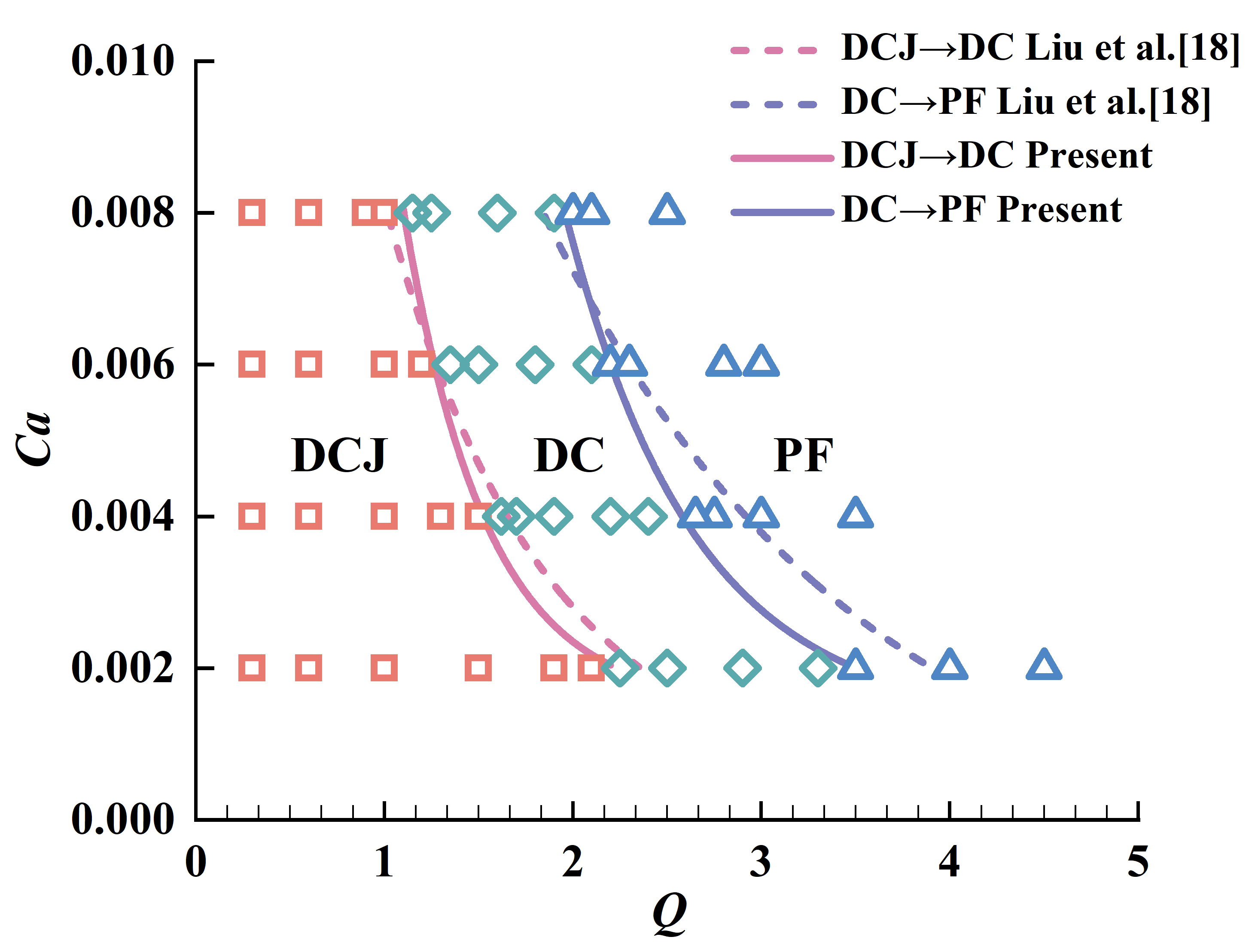}
  \caption{}
  \label{fig:2c}
  \end{subfigure}
    % 这里是整个大图的标题，编号会自动设为"图2"
  
  \captionsetup{justification=raggedright,singlelinecheck=false}
  \caption{Validation of grid independence and comparison of simulation results. (a) Grid distribution of Mesh4 in the micro-droplet channel. The fluid region is divided into triangular elements. The enlarged plot shows local grid refinement near the cell. The red dot in the middle of the junction is the monitoring point $P_M$. (b) The $x$-direction velocity of the monitoring point $P_M$ during $t$ = 18.0\textasciitilde{}21.0 ms under different numbers of grid elements. (c) The diagram of flow regimes during droplet generation with the flow rate ratio $Q$ and the capillary number $Ca$. The pink dotted line and the purple dotted line represent the transitional boundary between the flow regimes DCJ and DC, DC and PF, respectively, as obtained by Liu et al.18. The pink solid line and the purple solid line are our corresponding results. The red triangles, green diamonds, and blue squares represent our simulation results corresponding to DCJ, DC, and PF regimes.}
 
  \label{fig:2}
\end{figure*}  

\begin{table*}[ht]
\centering
\caption{Grid independence verification}

\label{tab:2}
% 将表格宽度设置为页面文本宽度
\begin{tabular*}{\textwidth}{@{\extracolsep{\fill}} c c c c c @{}}
\hline
\textbf{Grid} & \textbf{Number of Elements} & \textbf{ $\bm{V_{cell}}$ at t = 9.1 ms}& \textbf{Relative Error of $\bm{V_{cell}}$} & \textbf{Relative Error of Velocity at $P_M$} \\
\hline
  Mesh1 & 4.3w  & 12.91mm/s & 11.8\% & 13.8\% \\
  Mesh2 & 6.2w  & 13.31mm/s & 8.9\%  & 9.6\% \\
  Mesh3 & 8.1w  & 13.79mm/s & 5.7\%  & 6.1\% \\
  Mesh4 & 10.2w & 14.14mm/s & 3.4\%  & 4.6\%\\
  Mesh5 & 15.5w & 14.62mm/s & /      & /\\
\hline
\end{tabular*}
\end{table*}

The cell blockage ratio, ranging from 0 to 1, is defined by the dimensionless parameter $\Gamma=2R/W_{d}$, where $R$ is the cell radius. We set the cell density $\rho_{cell}$ equal to the dispersed phase density and ignore the effects of buoyancy or gravity. The shear stiffness factor $C_{1}$ in the M-R model reflects the elasticity of the cells. A larger $C_{1}$ corresponds to a stronger ability to resist deformation. The regulating factor $C_{2}$ describes the nonlinear mechanical properties of the cell under large deformation. When $C_{2} > 0$, the material exhibits "strain hardening" (the greater the deformation, the higher the stiffness); when $C_{2} < 0$, it exhibits "strain softening" (the greater the deformation, the lower the stiffness).

The dispersed phase is water at room temperature and normal pressure \cite{Raj2010}, and a surfactant is added to reduce the interfacial tension to 0.06 N/m \cite{Yaghoobi2020}.
The continuous phase is perfluorobutylamine FC-40 oil, with density measured by Pu et al.\cite{Pu2024}. Using Atomic Force Microscopy (AFM) and Total Internal Reflection Fluorescence Microscopy (TIRFM), Mathur et al.\cite{Mathur2000} obtained the elastic modulus of umbilical vein endothelial cells to be 1.3$\sim$7.2 kPa. Assuming the cells are incompressible, the Young's modulus $E = 6(C_{1} + C_{2})$. If $C_{2}$ is fixed at 200 Pa, then $C_{1}$ is 17$\sim$1000 Pa. The above experimental results provide important data support for subsequent numerical simulations, offering a solid quantitative basis for accurately depicting the dynamic behavior and mechanical properties of cell microdroplets in microchannels.

\subsection{\label{sec:level2}Validation of Algorithm}

To ensure the accuracy of the calculation results, for the model of cell movement in microchannels, we conducted grid independence verification using five different grid resolutions. Statistical analyses are performed on the average $x$-direction velocity of the cell $V_{cell}$ (defined in eq. (\ref{Vcell})) and the $x$-direction velocity at the monitoring point in the flow field for different meshes. The model parameters used for the grid independence verification are, $Ca$ = 0.004, $Re$ = 0.064, $Q$ = 1, $\Lambda$=1, $\lambda$=0.1, $L_{1}$ = 200 $\mu$m, $L_{2}$ = 1000 $\mu$m, and $D_{c}$ = 60 $\mu$m. The cell parameters are $C_{1}$ = 700 Pa, $C_{2}$ = 200 Pa, $\Gamma$ = 0.30. The results of validation are shown in Figure 2(b).

The grid is locally refined in the fluid-structure interaction regions to capture interface details. To verify grid convergence, we compare the deviations in cell velocity and monitoring point values relative to the reference solution (Mesh 5). Table II demonstrates that Mesh 4 (102,000 elements) achieves grid-independent results within an acceptable resolution. Therefore, Mesh 4 is adopted to ensure a balance between numerical accuracy and computational efficiency. 
  
To further validate the numerical framework, the simulated flow regime map is compared with the 3D lattice Boltzmann results of Liu et al.\cite{Liu2011}, as illustrated in Fig. \ref{fig:2c}. The current 2D results show excellent agreement with the reference data. The transition boundary between the flow regimes demonstrates an inverse correlation where $Ca$ decreases as $Q$ increases. Physically, this trend reflects the force balance, i.e., at higher flow rate ratios $Q$, the enhanced momentum of the dispersed phase facilitates breakup, thereby requiring less viscous stress (lower $Ca$) from the continuous phase to trigger the topological transition. Quantitative analysis yields an average relative error of 7.8\%, confirming the accuracy of the 2D approximation in capturing the critical encapsulation conditions. Therefore, the numerical model is reliable for the subsequent investigation of the DCJ and DC regimes.

\section{\label{sec:level2}RESULTS AND DISCUSSION}

\subsection{\label{sec:level2} Cell Dynamics in Droplet Encapsulation}

Considering the prevalence of the DCJ mode in cell encapsulation applications, we begin by examining the hydrodynamics in cell-laden droplet generation. The parameters for this simulation are $Ca$ = 0.004, $Re$ = 0.064, $Q$ = 1, $\Lambda$=1, $\lambda$=0.1, $L_{1}$ = 200 $\mu$m, $L_{2}$ = 1000 $\mu$m, and $D_{c}$ = 60 $\mu$m.  The cell parameters are $C_{1}$ = 700 Pa, $C_{2}$ = 200 Pa, and $\Gamma$ = 0.30. By observing the movement and deformation of the cell over time (Figure 3), we analyzed the encapsulation process in microdroplets.

To quantify the mechanical response of the cell during droplet generation, the von Mises stress\cite{Wang1997} is calculated in the present study. 
According to the stress invariant theory, the von Mises stress is obtained from the principal stresses as,
  \begin{equation}
  \sigma_{vM} = \sqrt{\frac{1}{2}[(\sigma_{1}-\sigma_{2})^{2} +(\sigma_{2}-\sigma_{3})^{2}+ (\sigma_{3}-\sigma_{1})^{2}]},
  \end{equation}
where $\sigma_{i}$ ($i$ = 1, 2, 3) denote the principal stresses obtained from the solid mechanics solver.

As illustrated in Figure 3, the temporal evolution of the cell encapsulation process can be divided into four stages.

\noindent{\bf 1. Steady movement} 

Given that the cell diameter is comparable to the microchannel width, it experiences a spatially inhomogeneous flow field. Following the approach of White et al. \cite{White2006}, the characteristic velocity when the cell travels in the channel is defined as the average of the maximum and minimum velocities, i.e., $V_{\mathrm{cell}} =(V_{\mathrm{max}} + V_{\mathrm{min}})/2$. $V_\mathrm{max}$ represents the velocity at the center of the cell,
  \begin{equation}
  V_{\mathrm{max}} = A_{0} \frac{Q_{d}}{W_{d}^{2}},
  \end{equation}
where $A_{0}$ is an empirical coefficient. If the flow is simplified to one-dimensional flat plate flow, the ratio of the maximum velocity to the average velocity is $A_{0}$ = 1.5.

According to the velocity distribution for Poiseuille flow, the edge velocity $V_{\mathrm{min}}$ of the cell can be estimated by\cite{White2006},
  \begin{equation}
  V_{\mathrm{min}} = V_{\mathrm{max}}(1-\Gamma^{2}).
  \end{equation}
Here, $\Gamma$ is the cell blockage ratio. This yields the average cell velocity $V_{\mathrm{cell}}$,
  \begin{equation}\label{Vcell}
  V_{\mathrm{cell}} = A_{0} \frac{2 - \Gamma^{2}}{2W_{d}^{2}} Q_{d}.
  \end{equation}
  
\begin{figure*}[ht]
    \centering
    % 主图 (a)
    \setcounter{subfigure}{0}
    \begin{subfigure}[b]{0.32\textwidth}
        \centering
        \includegraphics[width=\textwidth]{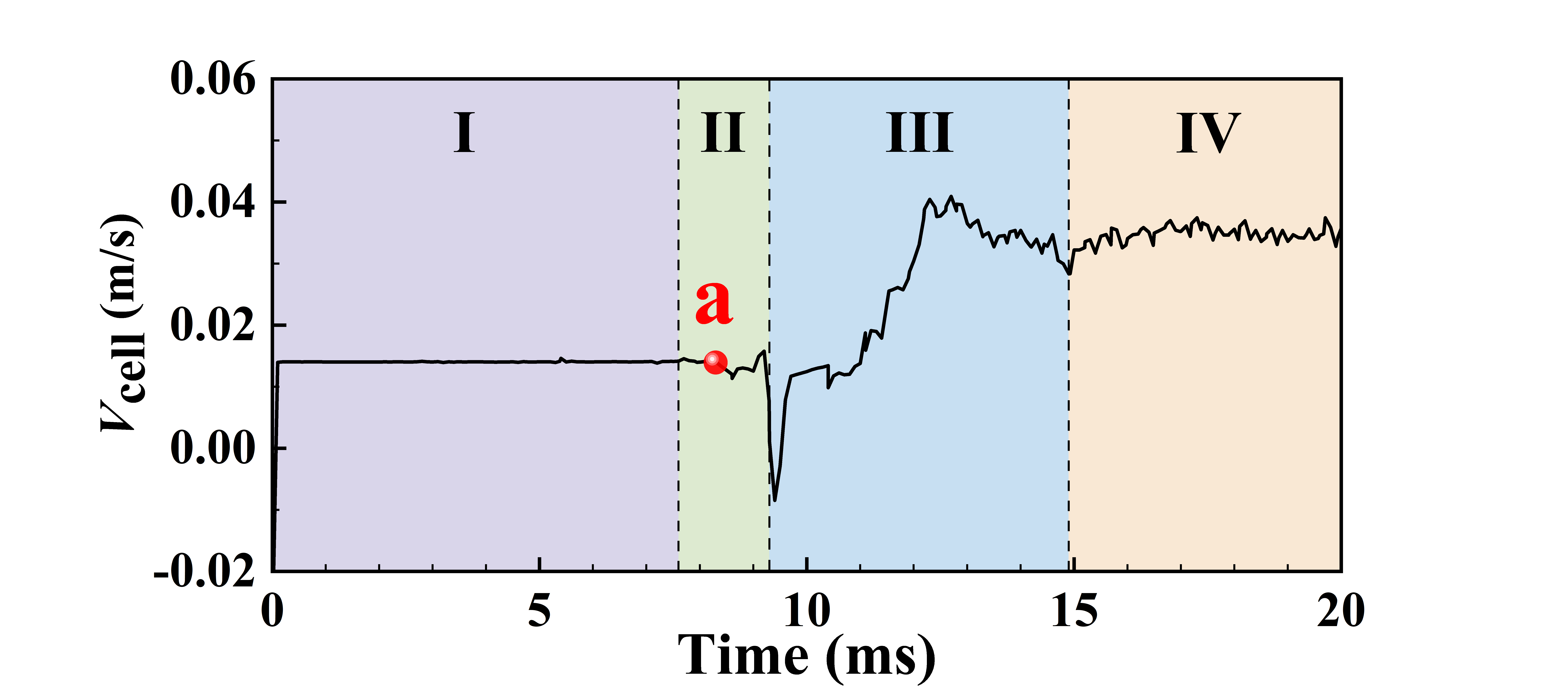}
        \captionsetup{justification=raggedright,singlelinecheck=false}
        \caption{\hskip 2cm $t=$ 8.3 ms}
        \label{fig:3a}
    \end{subfigure}
    \hfill
    % 子图 a(i)
    \begin{subfigure}[b]{0.2\textwidth}
        \centering
        \includegraphics[width=\textwidth]{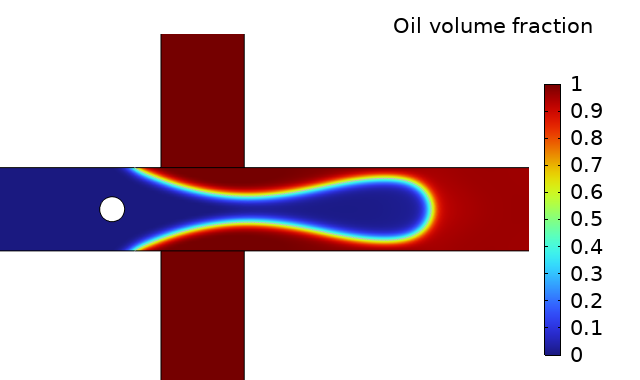}
        % 临时重定义子图编号
        \captionsetup[subfigure]{aboveskip=0pt,belowskip=0pt}
        \renewcommand{\thesubfigure}{i}
        \caption{}
        \label{fig:3a(i)}
    \end{subfigure}
    \hfill
    % 子图 a(ii)
    \begin{subfigure}[b]{0.2\textwidth}
        \centering
        \includegraphics[width=\textwidth]{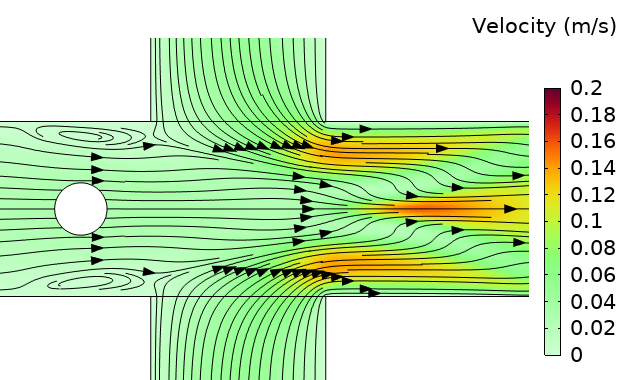}
        \renewcommand{\thesubfigure}{ii}
        \caption{}
        \label{fig:3a(ii)}
    \end{subfigure}
    \hfill
    % 子图 a(iii)
    \begin{subfigure}[b]{0.2\textwidth}
        \centering
        \includegraphics[width=\textwidth]{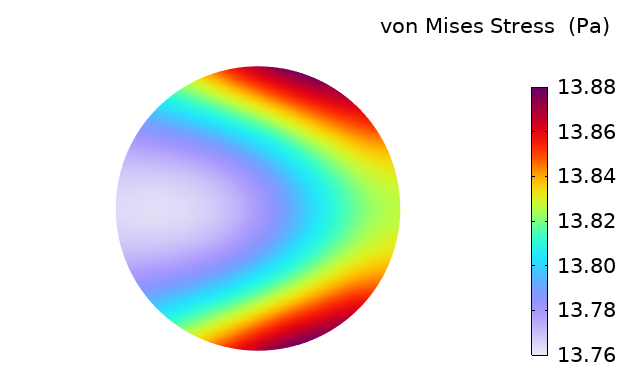}
        \renewcommand{\thesubfigure}{iii}
        \caption{}
        \label{fig:3a(iii)}
    \end{subfigure}
% 第二行：t=9.1ms
     \begin{subfigure}{0.32\textwidth}
     \setcounter{subfigure}{1}
        \centering
        \includegraphics[width=\textwidth]{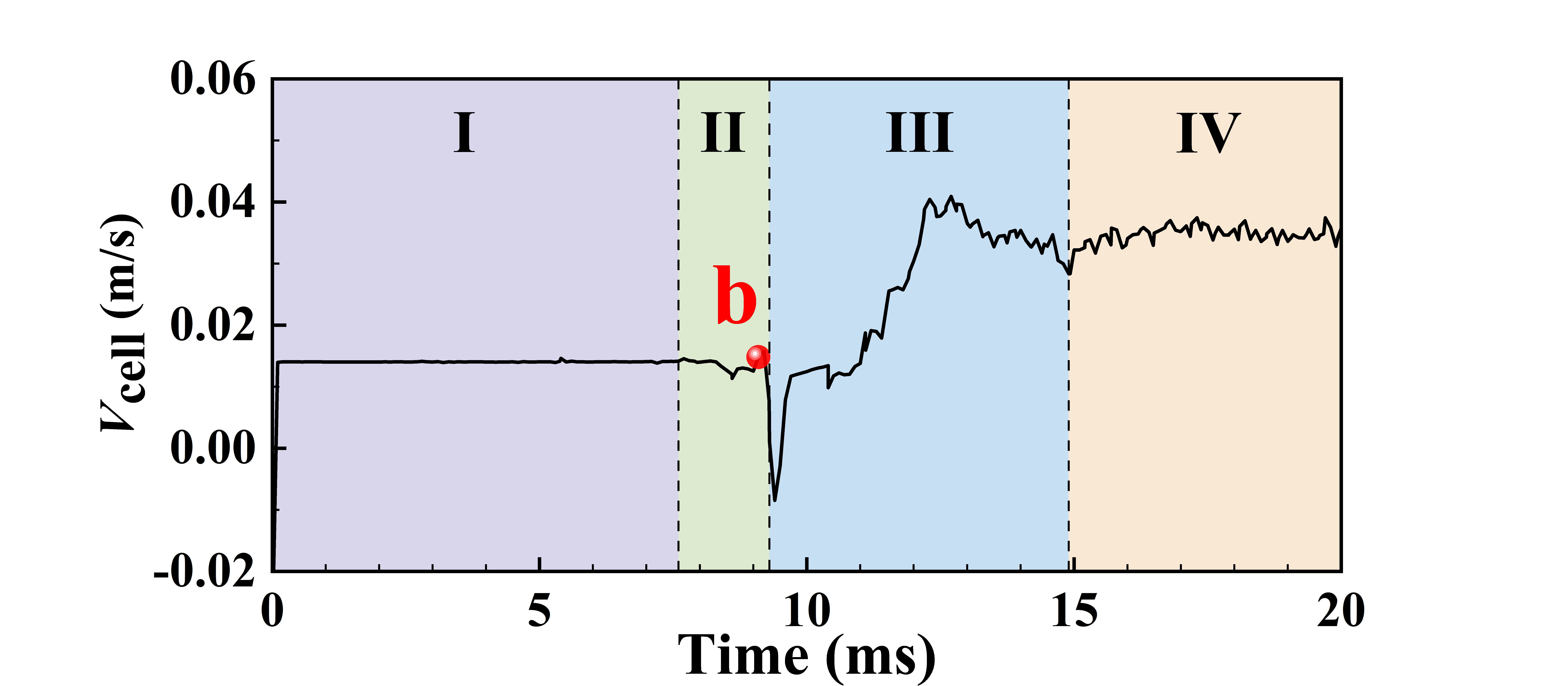}
        \captionsetup{justification=raggedright,singlelinecheck=false}
        \caption{\hskip 2cm $t=$ 9.1 ms}
        \label{fig:3b}
    \end{subfigure}
    \hfill % 子图之间的空白
    \begin{subfigure}{0.2\textwidth}
        \centering
        \includegraphics[width=\textwidth]{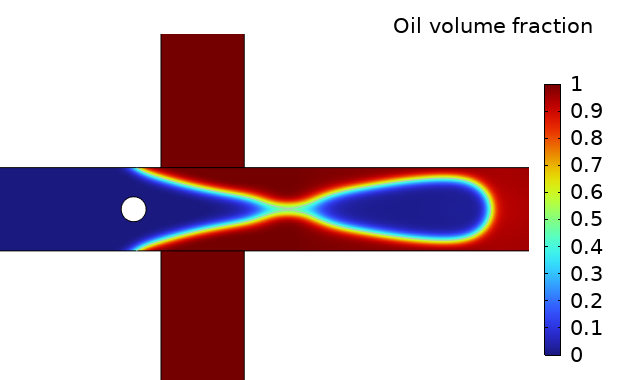}
         \renewcommand{\thesubfigure}{i}
        \caption{}
        \label{fig:3b(i)}
    \end{subfigure}
    \hfill
    \begin{subfigure}{0.2\textwidth}
        \centering
        \includegraphics[width=\textwidth]{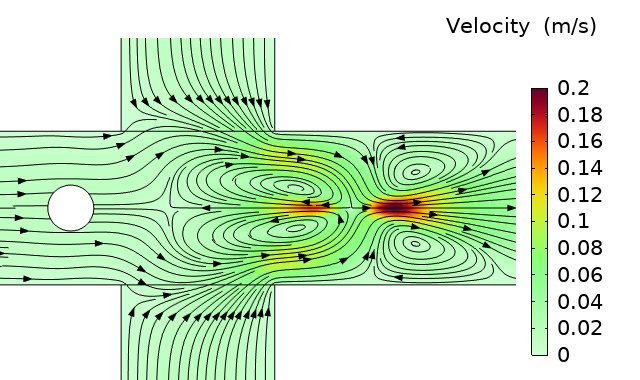}
        \renewcommand{\thesubfigure}{ii}
        \caption{}
        \label{fig:3b(ii)}
    \end{subfigure}
    \hfill
    \begin{subfigure}{0.2\textwidth}
        \centering
        \includegraphics[width=\textwidth]{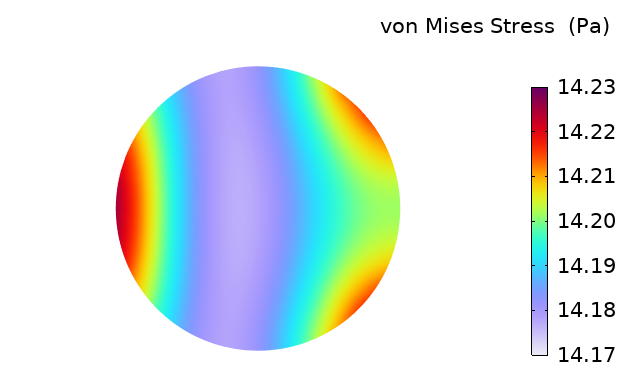}
       \renewcommand{\thesubfigure}{iii}
        \caption{}
        \label{fig:3b(iii)}
    \end{subfigure}
  
% 第三行：t=9.3ms
     \begin{subfigure}{0.32\textwidth}
     \setcounter{subfigure}{2}
        \centering
        \includegraphics[width=\textwidth]{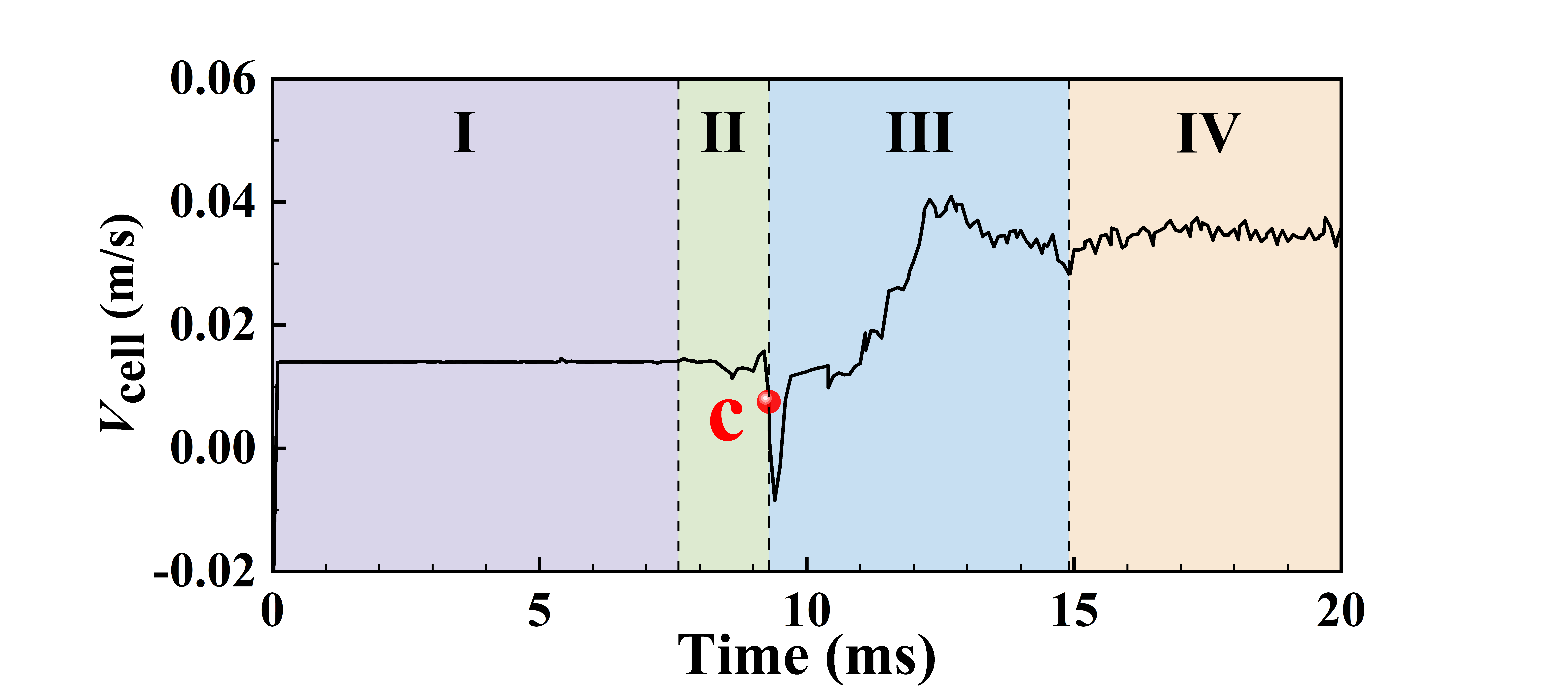}
        \captionsetup{justification=raggedright,singlelinecheck=false}
        \caption{\hskip 2cm $t=$ 9.3 ms}
        \label{fig:3c}
    \end{subfigure}
    \hfill % 子图之间的空白
    \begin{subfigure}{0.2\textwidth}
        \centering
        \includegraphics[width=\textwidth]{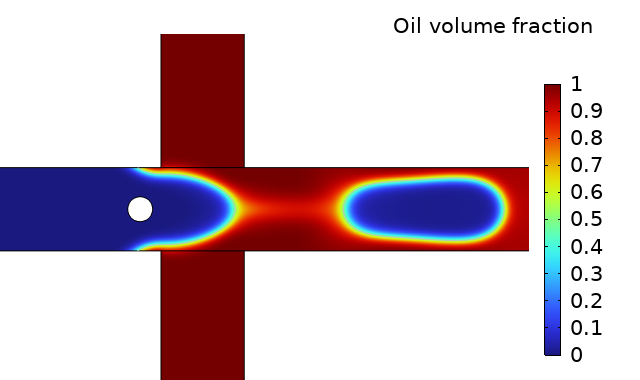}
       \renewcommand{\thesubfigure}{i}
        \caption{}
        \label{fig:3c(i)}
    \end{subfigure}
    \hfill % 子图之间的空白
    \begin{subfigure}{0.2\textwidth}
        \centering
        \includegraphics[width=\textwidth]{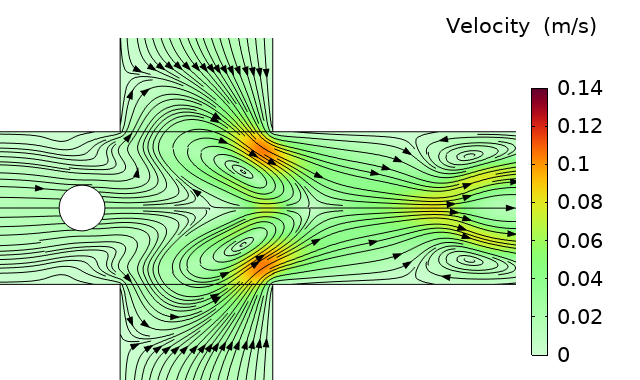}
        \renewcommand{\thesubfigure}{ii}
        \caption{}
        \label{fig:3c(ii)}
    \end{subfigure}
    \hfill
    \begin{subfigure}{0.2\textwidth}
        \centering
        \includegraphics[width=\textwidth]{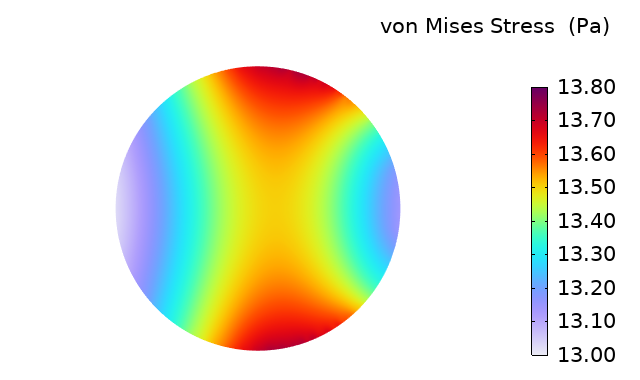}
        \renewcommand{\thesubfigure}{iii}
        \caption{}
        \label{fig:3c(iii)}
    \end{subfigure}
    
% 第四行：t=9.8ms
     \begin{subfigure}{0.32\textwidth}
     \setcounter{subfigure}{3}
        \centering
        \includegraphics[width=\textwidth]{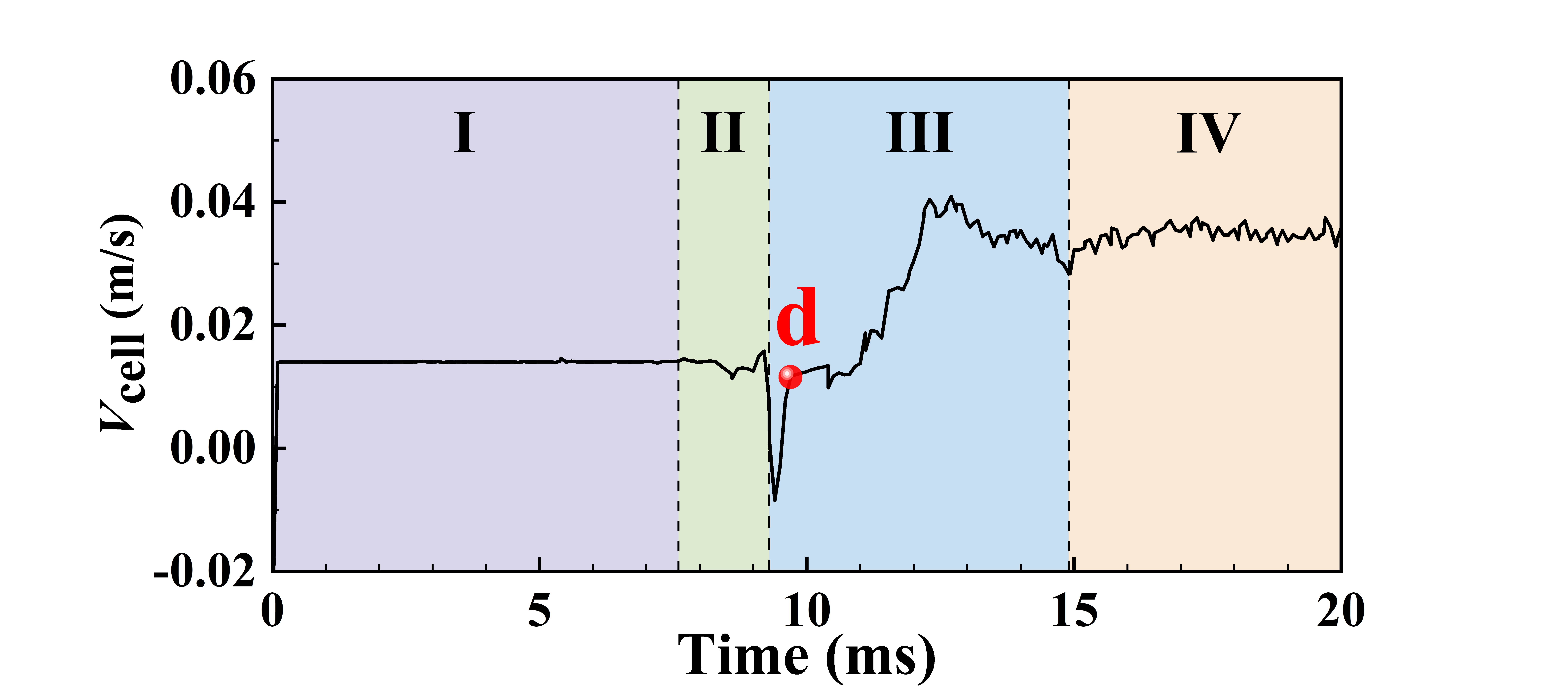}
        \captionsetup{justification=raggedright,singlelinecheck=false}
        \caption{\hskip 2cm $t=$ 9.8 ms}
        \label{fig:3d}
    \end{subfigure}
    \hfill % 子图之间的空白
    \begin{subfigure}{0.2\textwidth}
        \centering
        \includegraphics[width=\textwidth]{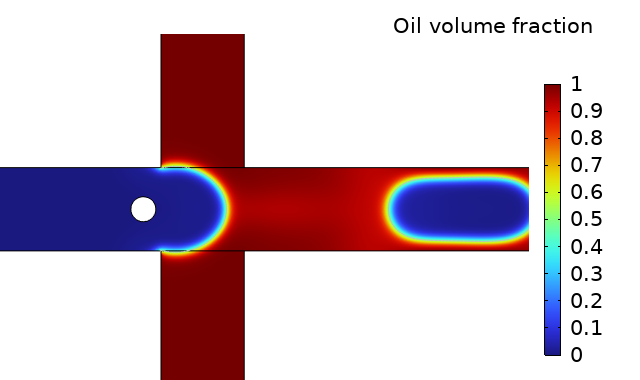}
         \renewcommand{\thesubfigure}{i}
        \caption{}
        \label{fig:3d(i)}
    \end{subfigure}
    \hfill
    \begin{subfigure}{0.2\textwidth}
        \centering
        \includegraphics[width=\textwidth]{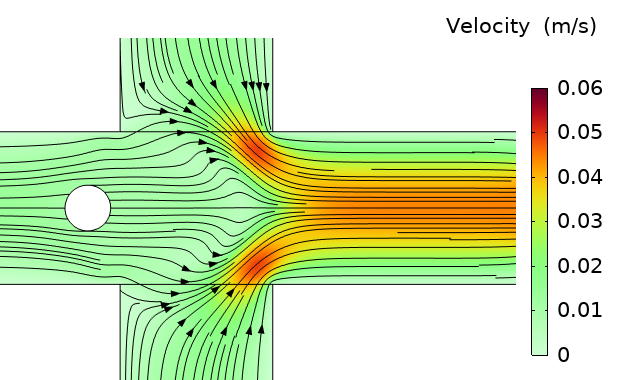}
         \renewcommand{\thesubfigure}{ii}
        \caption{}
        \label{fig:3d(ii)}
    \end{subfigure}
    \hfill
    \begin{subfigure}{0.2\textwidth}
        \centering
        \includegraphics[width=\textwidth]{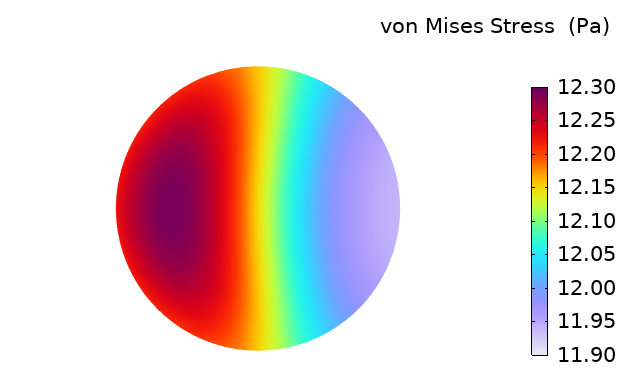}
         \renewcommand{\thesubfigure}{iii}
        \caption{}
        \label{fig:3d(iii)}
    \end{subfigure}
% 第五行：t=12.7ms
   \begin{subfigure}{0.32\textwidth}
   \setcounter{subfigure}{4}
        \centering
        \includegraphics[width=\textwidth]{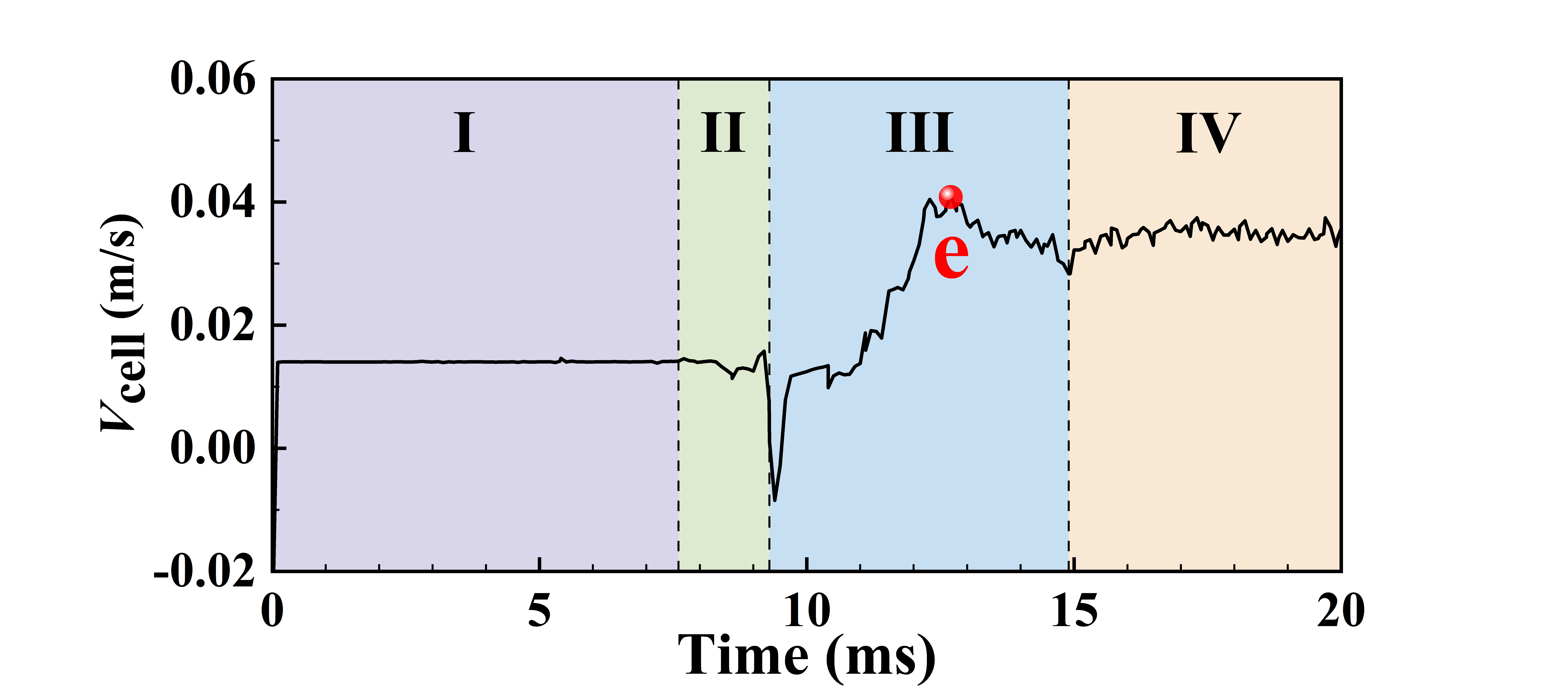}
        \captionsetup{justification=raggedright,singlelinecheck=false}
        \caption{\hskip 2cm $t=$ 12.7 ms}
        \label{fig:3e}
    \end{subfigure}
    \hfill % 子图之间的空白
  \begin{subfigure}{0.2\textwidth}
        \centering
        \includegraphics[width=\textwidth]{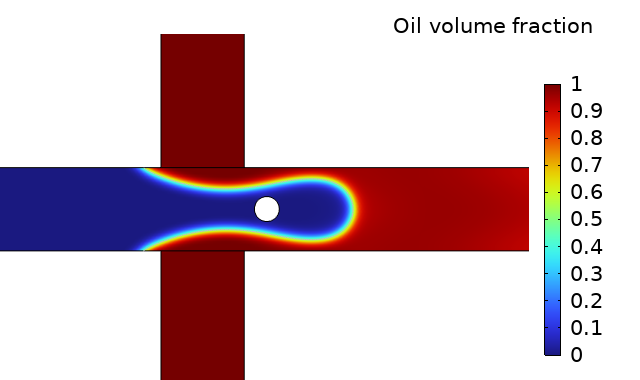}
         \renewcommand{\thesubfigure}{i}
        \caption{}
        \label{fig:3e(i)}
    \end{subfigure}
    \hfill
    \begin{subfigure}{0.2\textwidth}
        \centering
        \includegraphics[width=\textwidth]{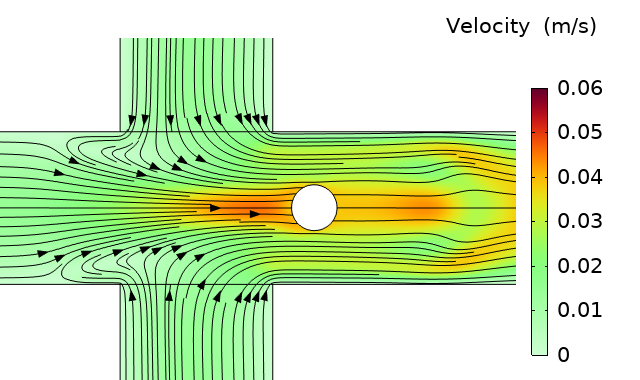}
        \renewcommand{\thesubfigure}{ii}
        \caption{}
        \label{fig:3e(ii)}
    \end{subfigure}
    \hfill
    \begin{subfigure}{0.2\textwidth}
        \centering
        \includegraphics[width=\textwidth]{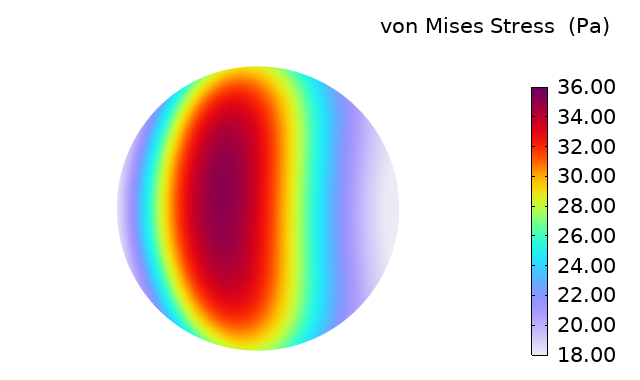}
         \renewcommand{\thesubfigure}{iii}
        \caption{}
        \label{fig:3e(iii)}
    \end{subfigure}
% 第六行：t=14.9ms
     \begin{subfigure}{0.32\textwidth}
     \setcounter{subfigure}{5}
        \centering
        \includegraphics[width=\textwidth]{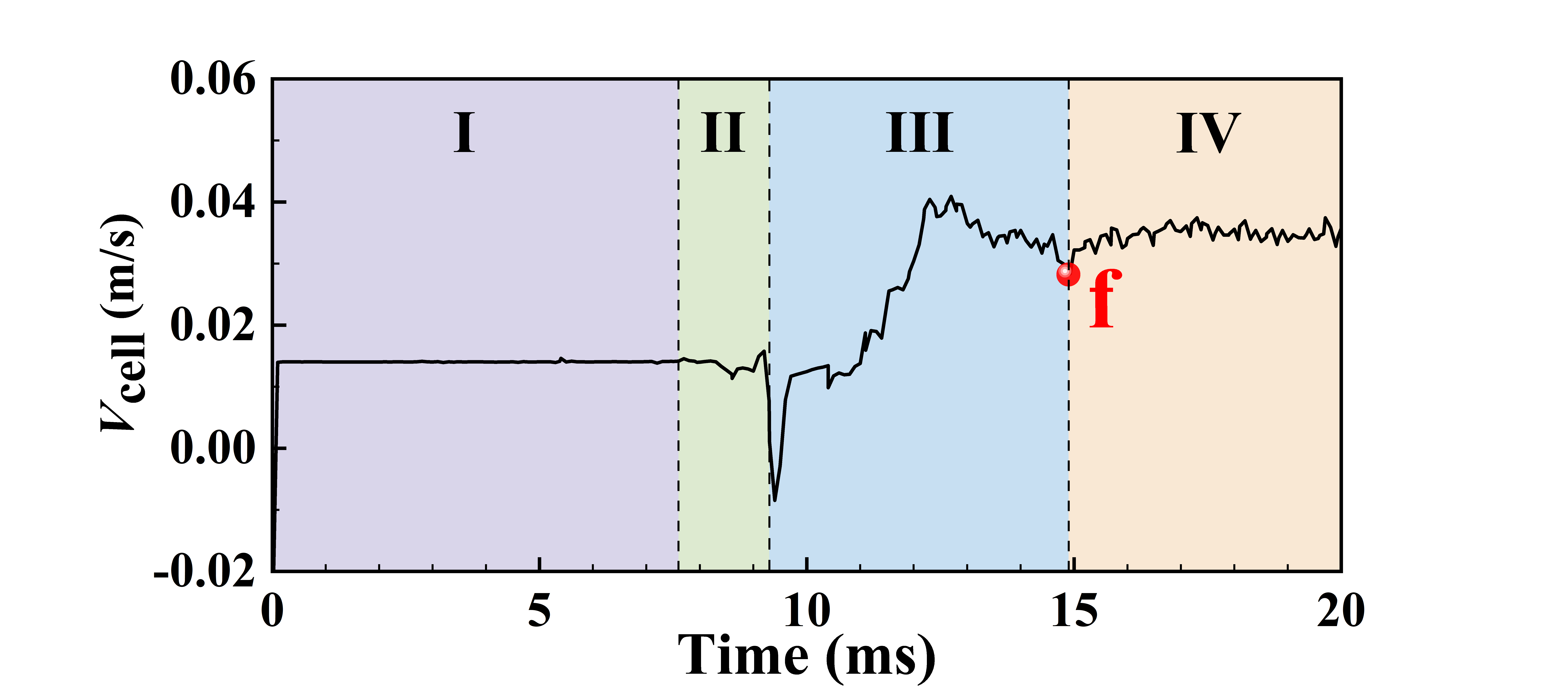}
        \captionsetup{justification=raggedright,singlelinecheck=false}
        \caption{\hskip 2cm $t=$ 14.9 ms}
        \label{fig:3f}
    \end{subfigure}
    \hfill % 子图之间的空白
    \begin{subfigure}{0.2\textwidth}
        \centering
        \includegraphics[width=\textwidth]{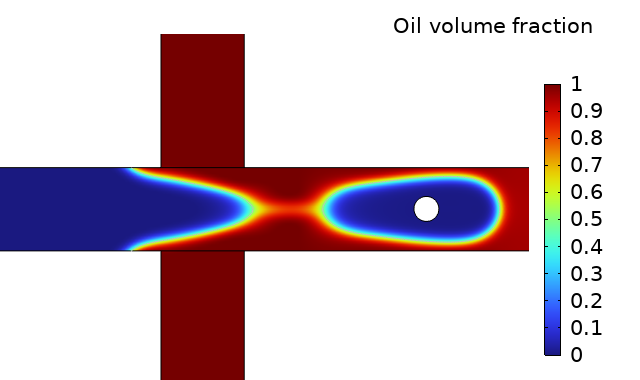}
         \renewcommand{\thesubfigure}{i}
        \caption{}
        \label{fig:3f(i)}
    \end{subfigure}
    \hfill 
    \begin{subfigure}{0.2\textwidth}
        \centering
        \includegraphics[width=\textwidth]{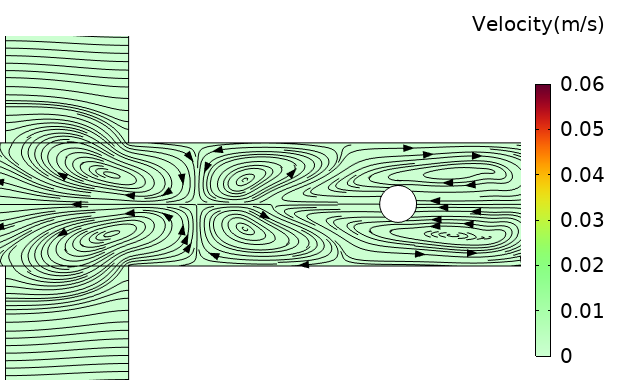}
         \renewcommand{\thesubfigure}{ii}
        \caption{}
        \label{fig:3f(ii)}
    \end{subfigure}
    \hfill
    \begin{subfigure}{0.2\textwidth}
        \centering
        \includegraphics[width=\textwidth]{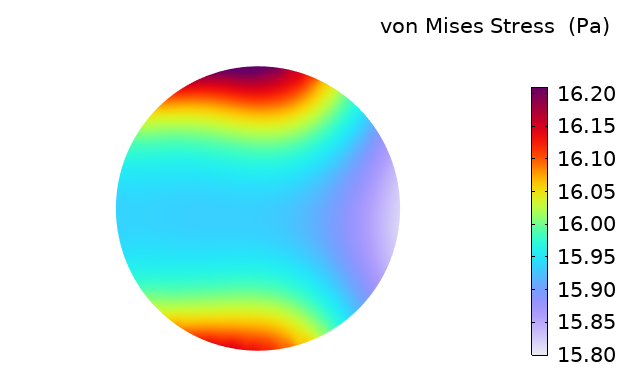}
         \renewcommand{\thesubfigure}{iii}
        \caption{}
        \label{fig:3f(iii)}
    \end{subfigure}
    \captionsetup{justification=raggedright,singlelinecheck=false}
    % 整个大组图的标题
    \caption{Fluid interface (i), flow field (ii), and stress distribution (iii) during cell encapsulation at different times. (a) $t$ = 8.3 ms; (b) $t$ = 9.1 ms; (c) $t$ = 9.3 ms; (d) $t$ = 9.8 ms; (e) $t$ = 12.7 ms; (f) $t$ = 14.9 ms. The left column illustrates the time history of cell speed $V_{cell}$. The parameters for this simulation are $Ca$ = 0.004, $Re$ = 0.064, $Q$ = 1, $\Lambda$=1, $\lambda$=0.1, $L_{1}$ = 200 $\mu$m, $L_{2}$ = 1000 $\mu$m, and $D_{c}$ = 60 $\mu$m. The cell parameters are $C_{1}$ = 700 Pa, $C_{2}$ = 200 Pa, and $\Gamma$ = 0.30.
    The color bar scales for velocity and cell stress vary across different time instants. In particular, at $t = 12.7$ ms, the stress magnitude experienced by the cell is significantly higher than at other stages.
    }
    \label{fig:3}
\end{figure*}

During the initial stage of cell movement ($t$ < 8.3 ms), the cells are carried by the dispersed phase and move steadily at the baseline speed within the channel due to the balance between fluid resistance and driving force, as shown in Fig. \ref{fig:3a} (Zone I). In this study, $A_{0}=1.47$ is measured, which deviates from the theoretical value by approximately 2\%. As we will discuss later, this steady velocity is crucial for successful encapsulation. 

\captionsetup[subfigure]{aboveskip=3pt,belowskip=1pt}

\noindent{\bf 2. Pre-encapsulation} 
  
As the cell continues to move forward, it is influenced by the prior droplet formation process, which is referred to as the pre-encapsulation stage (Points a and b in Fig. \ref{fig:3}(Zone II)). At $t$ = 8.3 ms, due to the previous droplet being in the neck contraction stage (Fig. 3a(i)), the velocity distribution within the channel evolves, and a backflow occurs in the corner area of the pipe intersection (Fig. 3a(ii)). The continuous phase fluid exerts a strong squeezing effect on the dispersed phase, and this squeezing stress is transmitted to the cell surface through the fluid medium. The head of the cell is then subjected to a large stress, as shown in Fig. 3a(iii). The cell experiences the fluid resistance and interface tension acting in a direction opposite to the cell movement, resulting in a slight decrease in $V_{cell}$.

Subsequently, the liquid bridge of the preceding droplet ruptures (Point b in Fig. \ref{fig:3b}, $t$ = 9.1 ms). As the neck contracts to a critical radius and pinches off (Fig. 3b(i)), the instantaneous release of interfacial energy triggers a violent hydrodynamic recoil, which induces backflow and vortex structures within the junction (see Fig. 3b(ii)). This vortex generation rapidly dissipates the radial squeezing pressure exerted by the continuous phase, thereby significantly reducing the blocking stress on the cell (Fig. 3b(iii)). Entrained by this vortex-induced flow, the cell velocity recovers rapidly to baseline levels. Meanwhile, the surrounding flow structure begins a significant reorganization. Streamlines converge from a regular laminar layer toward the junction center, while vorticity concentrates at the dispersed phase tip. This stage functions as a buffer period for energy accumulation and release, during which the cell completes passive response to pressure variations and prepares for the upcoming encapsulation process.
  
\noindent{\bf 3. Encapsulation} 

The encapsulation stage is the most critical period in the generation of cell-laden droplets (the corresponding time range is approximately $t=$ 9.3 $\sim$ 14.0 ms in Fig. \ref{fig:3}(Zone III)). This stage begins with intense interfacial retraction, followed by the filling of the dispersed phase, and finally, the droplet neck undergoes compression and rupture, completing the encapsulation of cells.  
 
At $t$ = 9.3 ms (Point c in Fig. \ref{fig:3c}), the interface curvature at the front of the dispersed phase reaches its maximum value after rupture of the previous liquid bridge. To minimize surface energy, the dispersed phase tip undergoes elastic retraction (Fig. 3c(i)). The streamline distribution shows that as the interface rapidly retreats, it induces a pair of strong counter-rotating recoil vortices at the tip of the dispersed phase fluid (Fig. 3c(ii)). This vortex is located near the central axis of the microchannel. Its rotational effect causes reverse flow, i.e., the fluid flows axially towards the entrance of the microchannel. The emergence of this reverse vortex structure decelerates cell movement, resulting in a significant reduction in cell speed over a short timeframe. Furthermore, the fluid close to the wall maintains forward movement simultaneously, which creates strong shear imposed on the sides of the cell, as illustrated in the stress distribution in Fig. 3c(iii).
It is worth noting that this flow-solid interaction, dominated by "retraction vortices," facilitates correcting the cell's lateral position, allowing it to align more precisely with the microchannel's central axis and thereby aiding in adjusting cell position during the encapsulation process.

As the interface retracts and the energy is dissipated, the dispersed phase fluid, driven by a constant flow rate at the inlet, begins to advance toward the centre of the cross-junction, and the system enters the filling stage. We select $t$ = 9.8 ms (Point d in Fig. \ref{fig:3d}) as the representative moment of this stage. At this instant, the dispersed phase interface gradually bulges outward from its original concave shape, forming a continuously expanding droplet interface (Fig. 3d(i)). At the filling stage, flow velocity is highest along the channel axis, behaving like a fountain flow. After impinging on the interface front, the flow forms curved streamlines (Fig. 3d(ii)). Meanwhile, the cell is pushed into the core of this fountain flow. Due to the velocity difference between the central flow and the moving interface, the cell accelerates continuously under the action of fluid pressure. The stress contours also indicate that the force on the cell's posterior is significantly greater than that on the anterior (Fig. 3d(iii)). This net thrust effectively clarifies the mechanism behind the cell's acceleration.

When the expanding dispersed phase droplet completely occupies the main channel cross-section (Point e in Fig. \ref{fig:3e}, $t$ = 12.7 ms), the process enters the necking stage. At this instant, the continuous phase fluid cannot directly pass through the cross-junction and is forced to flow through the narrow gaps between the droplet and the channel wall (Fig. 3e(i)). This obstruction induces a sharp rise in upstream pressure, forcing the continuous phase to squeeze the neck of the dispersed phase perpendicularly to the flow direction. Consequently, the flow field demonstrates strong extensional flow characteristics. Streamlines are highly contracted in the neck region, while the droplet interior demonstrates plug flow behavior due to the constriction effect of the neck (Fig. 3e(ii)). As the neck diameter diminishes and approaches the cell diameter, the cell induces a significant blockage effect on the flow field. Although the Bernoulli principle suggests that fluid velocity should increase through a contracted cross-section, the presence of the cell increases the local hydraulic resistance within this low-Reynolds-number, high-confinement regime. The cell becomes entrapped in the neck, restricting the supply of dispersed phase fluid, which results in a deceleration of both the cell and the surrounding fluid. Crucially, the cell is subjected to significant compressive stress from the continuous phase (Fig. 3e(iii)). This stage presents the highest risk of mechanical damage to the cell and serves as a critical screening window that determines the single-cell versus multi-cell encapsulation rates. Only if the cell successfully withstands this high stress can the droplet complete the pinch-off and achieve encapsulation.

\noindent{\bf 4. Post-Encapsulation} 

Upon the final rupture of the neck (pinch-off, $t > 14.9$ ms), the cell-laden droplet is generated and transported downstream(Fig. \ref{fig:3}(Zone IV)). Initially, the droplet assumes a stretched teardrop geometry (Fig. 3f(i)). Driven by interfacial tension, it rapidly recoils and relaxes into a spherical or plug-like configuration to minimize surface energy. This relaxation process induces transient internal mixing. Subsequently, as the droplet achieves steady transport driven by the continuous phase, the relative motion and shear interaction between the droplet interface and the channel wall (Fig. 3f(ii)) generate a pair of stable, centrosymmetric toroidal vortices. This phenomenon corresponds to the classic Hadamard-Rybczynski (H-R) circulation. Within this regime, fluid travels forward along the central axis and recirculates backward near the interface, establishing a complex flow topology around the solid core of the cell.

After encapsulation, the cell is no longer subject to the violent impact of external pressure gradients. Instead, its dynamics are governed by the droplet's internal circulation. As a suspended particle, the cell becomes entrained near the vortex axis. However, since the internal H-R circulation acts as a shear field rather than a rigid-body rotation, the dynamic coupling between the local velocity gradients and the cell's inertia and deformability leads to high-frequency, low-amplitude velocity fluctuations (Fig. \ref{fig:3}). Regarding the cell stress, it is found that the von Mises stress on the encapsulated cell (Fig. 3f(iii)) maintains a magnitude slightly higher than the pre-encapsulation baseline. This sustained stress arises from the interior pressure (Laplace pressure) within the droplet and the continuous viscous shear of the H-R vortex. Nevertheless, unlike the transient, significant stress variation observed during the necking stage, the stress in this stage is spatially uniform and temporally steady. This indicates that the cell has transitioned from a high-risk mechanical environment to a stable confinement state.

Employing a hyperelastic model, our analysis reveals the strong coupling between cell kinematics and droplet dynamics in the encapsulation process. Specifically, the impulsive release of interfacial energy during neck retraction triggers intense hydrodynamic recoil, subjecting the cell to peak mechanical stress, whereas the hydrodynamic driving force dominates the overall velocity evolution. Clarifying this interplay provides a theoretical foundation for optimizing flow parameters to minimize mechanical damage to the cell.

\subsection{\label{sec:level2}Critical condition for cell encapsulation}

\begin{figure*}[ht]
    \centering  % 只让图片居中\centering 
   
  \begin{subfigure}{0.55\textwidth}
  \centering 
  \includegraphics[width=\textwidth]{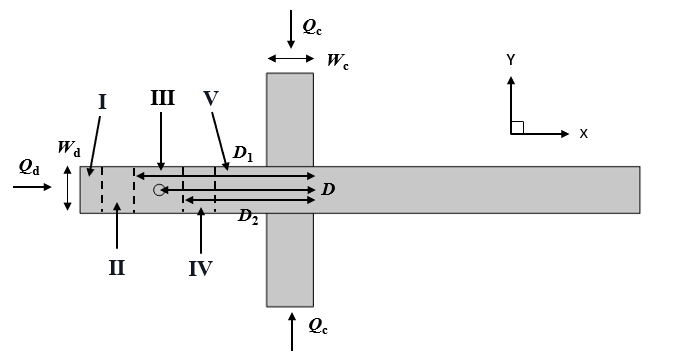}
  \caption{}
  \label{fig:4a}
  \end{subfigure}
  \hfill
  \begin{subfigure}{0.44\textwidth}
  \centering 
  \includegraphics[width=\textwidth]{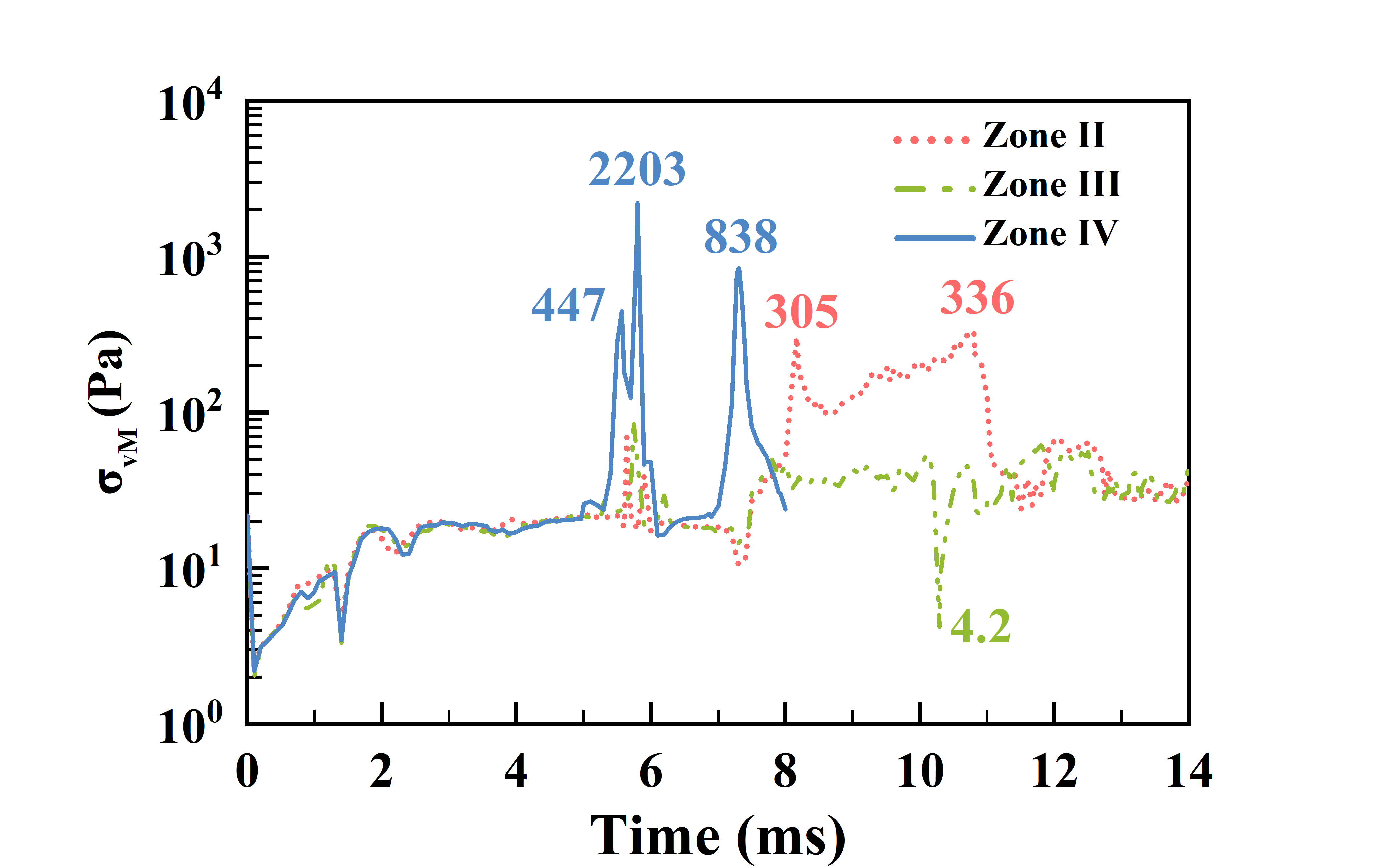}
  \caption{}
  \label{fig:4b}
  \end{subfigure}

  \captionsetup{justification=raggedright,singlelinecheck=false}
  \centering
  \caption{(a) Schematic diagram of the initial cell location for successful encapsulation.  Here, Zone I represents the advanced area before the encapsulation of the next droplet. Zone II represents the lagging area after the encapsulation of the current droplet. Zone III represents the normal encapsulation area of the current droplet. Zone IV represents the advanced area before the encapsulation of the current droplet. Zone V represents the lagging area after the encapsulation of the previous droplet. The distance from the left boundary of the successful encapsulation to the intersection of the microchannel is $D_{1}$ $\mu$m, and the distance from the right boundary of the successful encapsulation to the intersection is $D_{2}$ $\mu$m. (b) Von Mises stress vs. time curves of cells at different initial positions. The red dotted line corresponds to the cell initial position being in Zone II, the green double-dotted line in Zone III, and the blue solid line in Zone IV.
  }
  \label{fig:4}
\end{figure*}  

In microfluidic single-cell analysis technology, achieving efficient and deterministic single-cell encapsulation has always been a crucial challenge. Due to the random distribution characteristics of the cell suspension, the traditional droplet generation process often follows the Poisson distribution pattern, resulting in frequent occurrences of "empty droplets" or "multi-cell encapsulation" phenomena, and it is difficult to ensure the suitable positioning of cells within droplets. Moreover, if the cells do not enter the intersection junction at a phase matching the generation phase of the droplet, it not only leads to encapsulation failure but also may cause severe mechanical damage due to the high shear zone at the neck retraction stage. Therefore, clarifying the dependence of successful encapsulation on the initial position of the cell and establishing quantitative prediction criteria are crucial for optimizing chip design.

To explore the critical conditions for cell encapsulation, this study divides the main channel into specific regions (Fig. \ref{fig:4a}. Through numerical simulations, we track the dynamic trajectories and force responses of cells at different initial positions $D_c$ during the droplet generation process, thereby identifying the critical spatial window for normal encapsulation. Based on the phase difference relative to the droplet generation cycle, the initial cell positions are categorized into three typical regions: the Normal Encapsulation Mode (Zone III, with $D_{c}$ = 72 $\mu$m as an example), the Lagging Pinching Mode (Zone II, e.g., $D_{c}$ = 65 $\mu$m), and the Premature Escape Mode (Zone IV, e.g., $D_{c}$ = 77 $\mu$m). The simulation parameters are fixed at $Ca$ = 0.0018, $Re$ = 0.38, $Q$ = 0.5, $\Lambda$ = 1, $\lambda$ = 0.3, $L_{1}$ = 200 $\mu$m, and $L_{2}$ = 800 $\mu$m, with cell mechanical properties set to $C_{1}$ = 700 Pa, $C_{2}$ = 200 Pa, and $\Gamma$ = 0.40. Cells originating from these different regions exhibit divergent dynamic behaviors:

 \begin{figure*}[ht]
    \centering % 整个大组图在页面中居中
    \begin{subfigure}{0.31\textwidth}
        \centering
        \includegraphics[width=\textwidth]{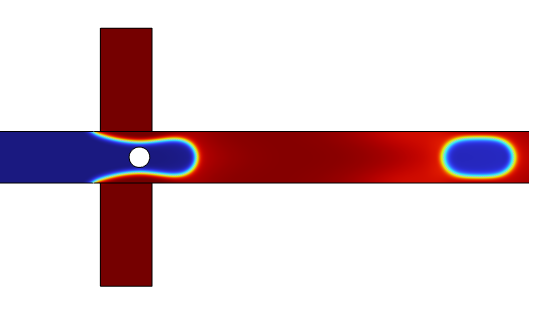}
        \caption{t=8.0ms}
        \label{fig:5a}
    \end{subfigure}
    \hfill % 子图之间的空白
    \begin{subfigure}{0.32\textwidth}
        \centering
        \includegraphics[width=\textwidth]{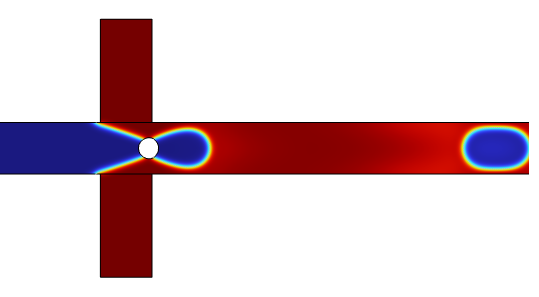}
        \caption{t=8.2ms}
        \label{fig:5b}
    \end{subfigure}
    \hfill
    \begin{subfigure}{0.33\textwidth}
        \centering
        \includegraphics[width=\textwidth]{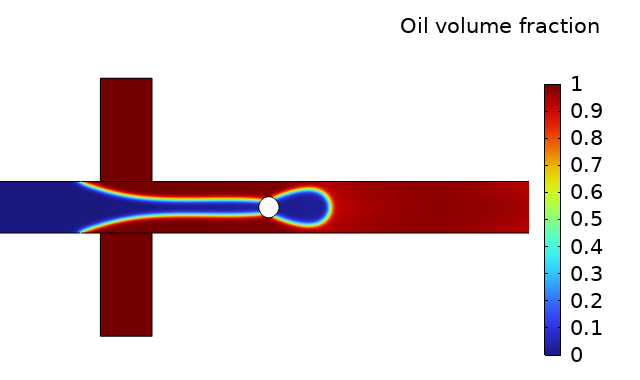}
        \caption{t=9.8ms}
        \label{fig:5c}
    \end{subfigure}

  % 整个大组图的标题
    \captionsetup{justification=raggedright,singlelinecheck=false}
    \caption{Flow interfaces at different instants for the cell initially located in Zone II. (a) $t$ = 8.0 ms, (b) $t$ =8.2 ms, and (c) $t$ =9.8 ms. The simulation parameters are fixed at $Ca$ = 0.0018, $Re$ = 0.38, $Q$ = 0.5, $\Lambda$ = 1, $\lambda$ = 0.3, $L_{1}$ = 200 $\mu$m, and $L_{2}$ = 800 $\mu$m, $D_{c}$ = 65 $\mu$m, with cell mechanical properties set to $C_{1}$ = 700 Pa, $C_{2}$ = 200 Pa, and $\Gamma$ = 0.40.}
    \label{fig:5}
\end{figure*}

{\bf Normal Encapsulation Mode (Zone III):} When the initial position of the cell falls within Zone III, the cell kinematics and the droplet generation cycle achieve a precise “spatiotemporal match”. The cell smoothly enters the droplet's bulk during the filling stage and is subsequently transported downstream. Throughout this process, the flow field provides an effective "buffer protection", exerting quasi-steady moderate forces on the cell without significant stress fluctuations (green curve in Fig. \ref{fig:4b}). Consequently, damage-free encapsulation is achieved. The detailed dynamics of this specific mode have been elaborated in the preceding section.
  
{\bf Lagging Pinching Mode (Zone II):} When the initial position of the cell falls within Zone II, its arrival at the junction lags behind the optimal window for droplet generation (Fig. \ref{fig:5}). At $t = 8.2$ ms, the cell is located precisely at the critical region of neck constriction (Fig. \ref{fig:5b}). Due to the cell's blocking effect, the local interfacial dynamics are disrupted. Consequently, the cell is directly subjected to strong radial compression from the continuous phase, resulting in a sharp increase in von Mises stress to 305 Pa. This "neck clamping" effect not only hinders the normal pinch-off of the droplet but may also induce severe deformation or even threaten the cell's viability.
  
\begin{figure*}[ht]
    \centering % 整个大组图在页面中居中 
    \begin{subfigure}{0.31\textwidth}
        \centering
        \includegraphics[width=\textwidth]{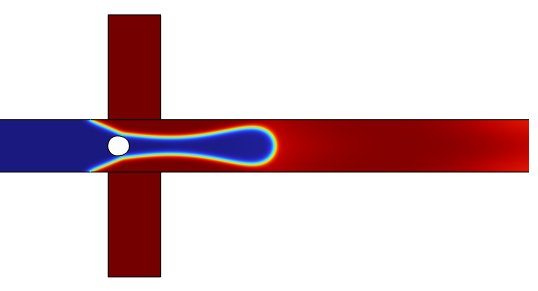}
        \caption{t=5.5ms}
        \label{fig:6a}
    \end{subfigure}
    \hfill % 子图之间的空白
    \begin{subfigure}{0.31\textwidth}
        \centering
        \includegraphics[width=\textwidth]{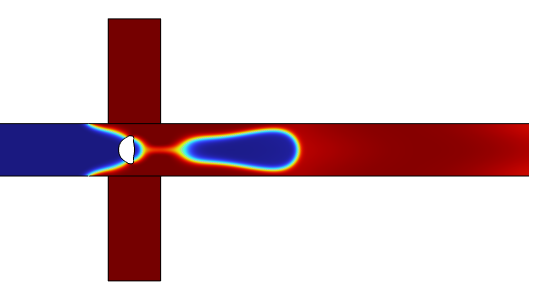}
        \caption{t=5.8ms}
        \label{fig:6b}
    \end{subfigure}
    \hfill
    \begin{subfigure}{0.33\textwidth}
        \centering
        \includegraphics[width=\textwidth]{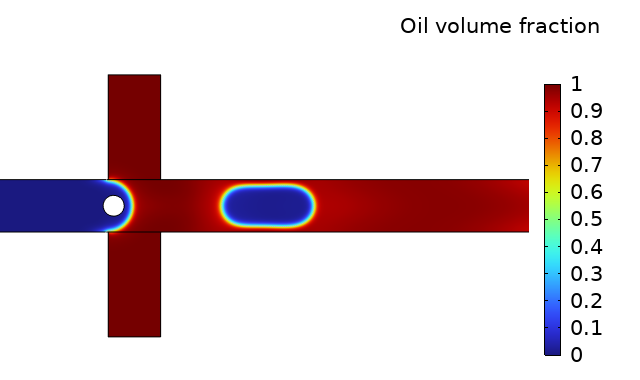}
        \caption{t=6.0ms}
        \label{fig:6c}
    \end{subfigure}
    \vspace{10pt} % 第一行和第二行之间的垂直间距

    \begin{subfigure}{0.315\textwidth}
        \centering
        \includegraphics[width=\textwidth]{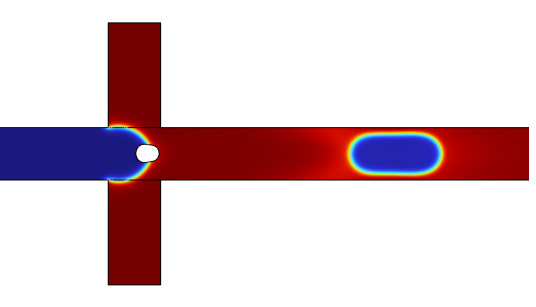}
        \caption{t=7.3ms}
        \label{fig:6d}
    \end{subfigure}
    \hfill
    \begin{subfigure}{0.32\textwidth}
        \centering
        \includegraphics[width=\textwidth]{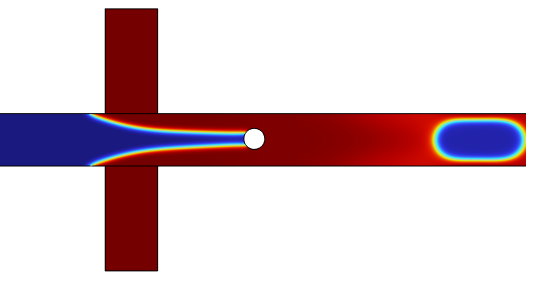}
        \caption{t=8.2ms}
        \label{fig:6e}
    \end{subfigure}
    \hfill
    \begin{subfigure}{0.34\textwidth}
        \centering
        \includegraphics[width=\textwidth]{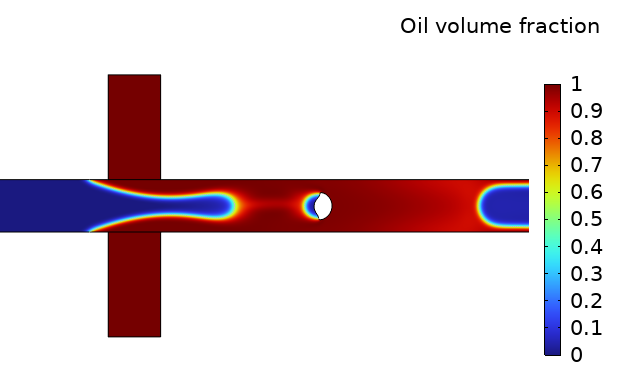}
        \caption{t=8.6ms}
        \label{fig:6f}
    \end{subfigure}
    \captionsetup{justification=raggedright,singlelinecheck=false}
    % 整个大组图的标题
    \caption{Flow interfaces at different instants for the cell initially located in Zone IV. (a) $t$ = 5.5 ms, (b) $t$ = 5.8 ms, (c) $t$ = 6.0 ms, (d) $t$ = 7.3 ms, (e) $t$ = 8.3 ms, and (f) $t$ = 8.6 ms. The simulation parameters are fixed at $Ca$ = 0.0018, $Re$ = 0.38, $Q$ = 0.5, $\Lambda$ = 1, $\lambda$ = 0.3, $L_{1}$ = 200 $\mu$m, and $L_{2}$ = 800 $\mu$m, $D_{c}$ = 77 $\mu$m, with cell mechanical properties set to $C_{1}$ = 700 Pa, $C_{2}$ = 200 Pa, and $\Gamma$ = 0.40.}
    \label{fig:6}
\end{figure*}

{\bf Premature Escape Mode (Zone IV):} When the cell is initially located within Zone IV, it demonstrates a characteristic "impact-recapture-breakthrough" dynamic sequence, accompanied by an extremely high risk of mechanical damage (Fig. \ref{fig:6}). At the moment of the previous droplet pinch-off ($t = 5.8$ ms), the intense hydrodynamic impulse generated by the interfacial recoil directly impacts the cell, subjecting it to a peak stress of 2203 Pa (Fig. \ref{fig:4b}, blue curve). Although the cell is subsequently recaptured by the dispersed phase, its velocity in the confined channel exceeds that of the bulk flow. Consequently, the cell penetrates the front interface again at $t = 7.3$ ms (breakthrough), becoming exposed to the continuous oil phase and enduring a secondary peak stress of 838 Pa. Ultimately, the cell prematurely escapes as it fails to be trapped by the forming droplet, leading to encapsulation failure.
  
To clarify the critical conditions for encapsulation dependent on the initial cell position, it is essential to consider the competition between the timescales of cell transport and droplet formation. The success of encapsulation is primarily determined by two key temporal parameters: the cell crossing time $T_{t}$, defined as the duration required for the cell to traverse the channel junction, and the droplet pinch-off time $T_{p}$, representing the interval from cycle initiation to neck rupture.

To systematically elucidate the regulatory mechanisms of the capillary number $Ca$ and flow rate ratio $Q$ on the cell encapsulation process, we develop a theoretical model to derive the scaling laws governing the critical spatial boundaries of the successful encapsulation window, denoted as $D_{1}$ and $D_{2}$, as shown in Fig. \ref{fig:4a}. 

The cell crossing time $T_{t}$ is defined as the duration required for the cell to traverse the characteristic distance $D$ (from its initial position to the junction) at its average velocity. Combining this with the cell velocity derived in Eq. \ref{Vcell}, the expression for $T_{t}$ is given by:
\begin{equation}\label{crossingtime}
T_{t} = \frac{D}{V_{\mathrm{cell}}} = \frac{2 D W_{d}^{2}}{A_{0}(2-\Gamma^{2}) Q_{d}},
\end{equation}
where $D$ represents the initial position (corresponding to the boundary values $D_1$ or $D_2$, as shown in Fig. \ref{fig:4a}).

The competition between viscous forces and interfacial tension dominates the droplet formation. Based on the scaling laws for T-junction droplet generation established by Garstecki et al.\cite{Garstecki2006} and refined by subsequent studies\cite{Xu2008, Liu2011}, the droplet break-up time $T_p$ generally follows a power-law relationship with $Ca$ and a linear dependence on $Q$, assuming mass conservation. Consequently, the model for $T_p$ can be expressed as:
\begin{equation}\label{dropletperiod}
T_{p}=\frac{V_{\mathrm{drop}}}{Q_{d}}=\frac{K \left(1+\alpha Q\right) W_{d}^{3} \mathrm{Ca}^{-n}}{Q_{d}}.
\end{equation}
Here, $K$ and $\alpha$ are geometric coefficients, and $n$ is the power-law exponent characterizing the pinch-off dynamics.

A successful encapsulation requires a precise spatiotemporal match between the cell's arrival and the droplet's pinch-off. The critical boundaries of the encapsulation window, $D_1$ and $D_2$, correspond to the limiting conditions where the cell transport time equals the droplet break-up time $T_t = T_p$. 

Equating Eqs. (\ref{crossingtime}) and (\ref{dropletperiod}), and solving for the critical position $D$, yields the general scaling law:
\begin{equation}
D_{\mathrm{critical}} = \frac{K A_{0} W_{d}\left(2-\Gamma^{2}\right)}{2}\left(1+\alpha Q\right) C a^{-n}.
\end{equation}
Since the encapsulation window is bounded by two distinct failure modes (Lagging Pinching at one end and Premature Escape at the other), the critical parameters $D_1$ and $D_2$ share the same functional form but possess distinct proportionality constants ($K_1, \alpha_1$ for $D_1$; $K_2, \alpha_2$ for $D_2$). Therefore, the criterion for successful single-cell encapsulation can be mathematically expressed as the inequality:
\begin{equation}
D_{2}(Ca, Q) \leq D_{\mathrm{c}} \leq D_{1}(Ca, Q).
\end{equation}
\begin{figure*}[]
    \centering  % 只让图片居中\centering 
  \begin{subfigure}{0.5\textwidth}
  \centering 
  \includegraphics[width=\textwidth]{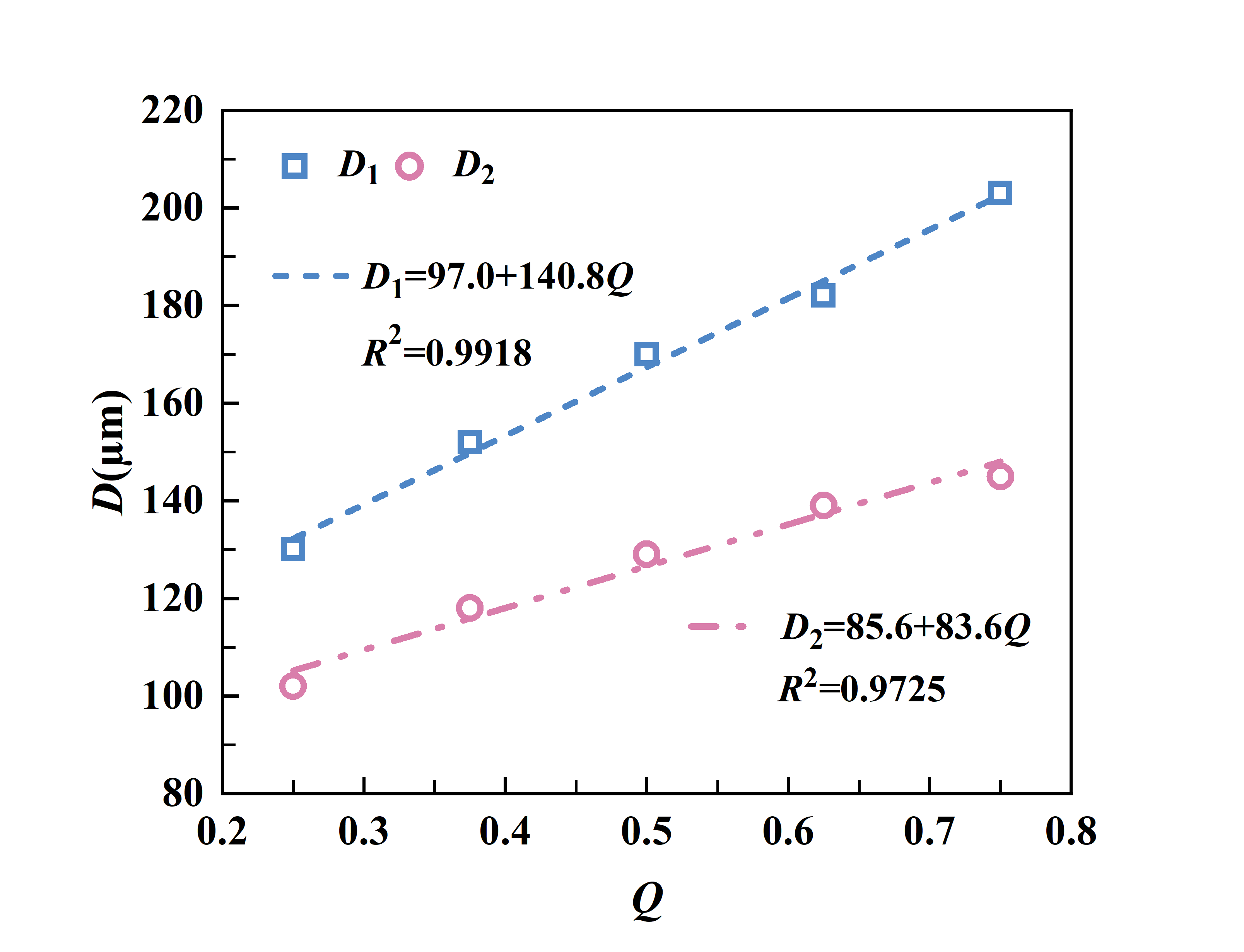}
  \caption{}
  \label{fig:7a}
  \end{subfigure}
  \hfill
  \begin{subfigure}{0.49\textwidth}
  \centering 
  \includegraphics[width=\textwidth]{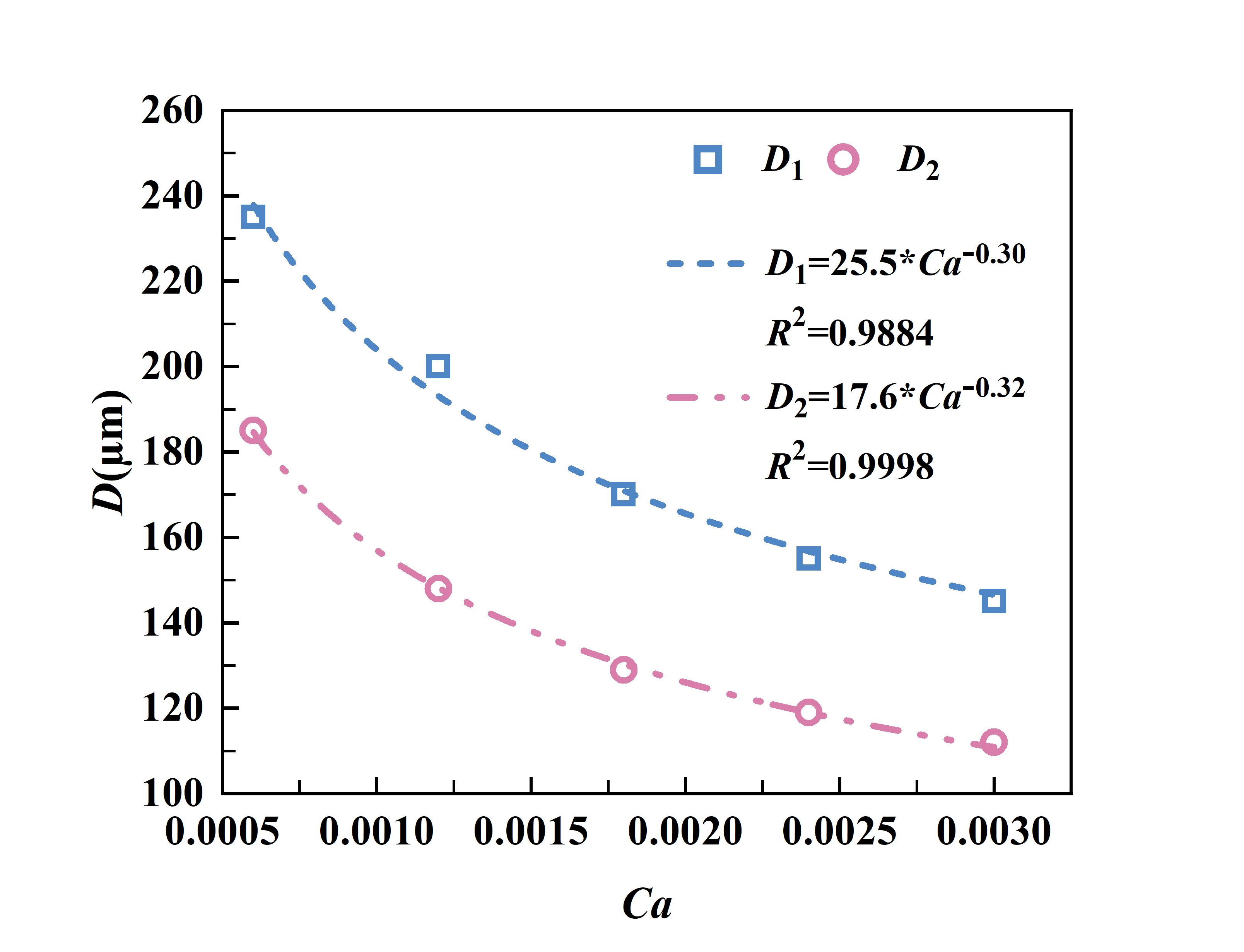}
  \caption{}
  \label{fig:7b}
  \end{subfigure}
  
  \captionsetup{justification=raggedright,singlelinecheck=false}
  \centering
  \caption{Comparison of theoretical and numerical results for encapsulation boundaries $D_1$ and $D_2$ dependent on $Q$ and $Ca$. (a) $D_1$ and $D_2$ vary linearly with $Q$; (b) $D_1$ and $D_2$ follow a power-law relationship with $Ca$. Simulation parameters are $Re$ = 0.38, $\Lambda$=1, $\lambda$=0.3, channel geometry $L_{1}$ = 300 $\mu$m, $L_{2}$ = 600 $\mu$m, and cell properties $C_{1}$ = 700 Pa, $C_{2}$ = 200 Pa, $\Gamma$ = 0.24.}
 \label{fig:7}
\end{figure*} 

To verify this model and determine the unknown constants, we conduct systematic numerical simulations with parameters at $Re$ = 0.38, $\Lambda$=1, $\lambda$=0.3, channel geometry $L_{1}$ = 300 $\mu$m, $L_{2}$ = 600 $\mu$m, and cell properties $C_{1}$ = 700 Pa, $C_{2}$ = 200 Pa, $\Gamma$ = 0.24. Our fitting results indicate that the power-law exponent $n$ is approximately 0.3 for both boundaries, which is consistent with previous experimental studies on droplet microfluidics. For instance, Christopher et al.\cite{Christopher2008} reported an exponent of 0.3 for droplet volume scaling, and Liu et al.\cite{Liu2011} observed an exponent range of 0.2 to 0.3. Fig. \ref{fig:7} depicts the comparison of theoretical and numerical results for encapsulation boundaries $D_1$ and $D_2$ dependent on $Q$ and $Ca$, where both boundaries scale linearly with $Q$ and follow a power-law relationship with $Ca$. The agreement between our derived scaling law and the simulation results confirms that the proposed model effectively captures the physical mechanism of cell encapsulation.

\subsection{\label{sec:level2}Phase Diagram of Cell-Laden Droplets}

\begin{figure}
    \captionsetup{justification=raggedright,singlelinecheck=false}
    \centering
    \includegraphics[width=1\linewidth]{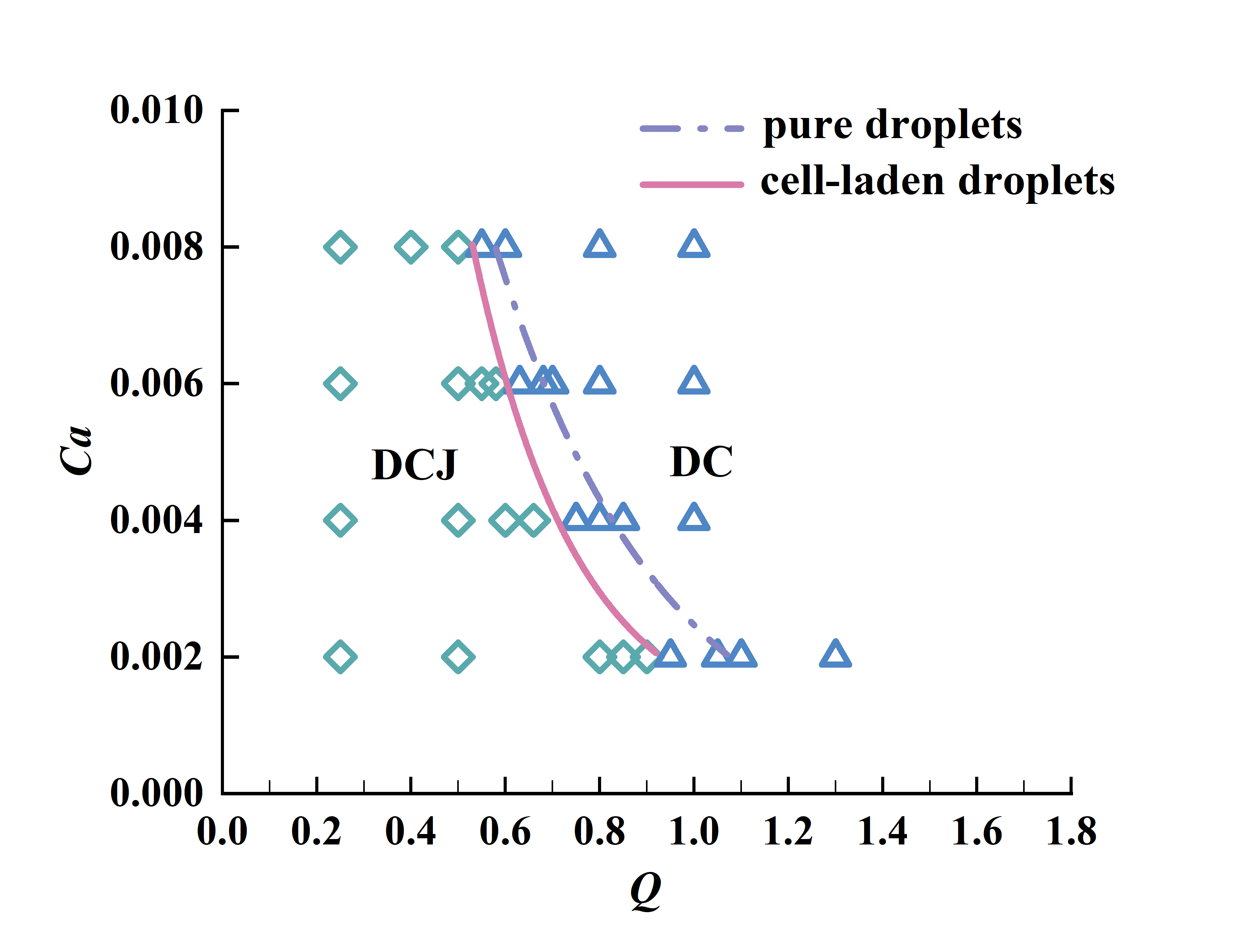}
    \caption{Phase diagram of cell-laden droplets with the flow rate ratio $Q$ and the capillary number $Ca$. The dashed line represents the transitional curve between DCJ and DC flow regimes for pure droplets, while the solid line represents cell-laden droplets. The green diamonds and blue triangles represent our simulation results corresponding to DCJ and DC regimes, respectively.
   }
    \label{fig:8}
\end{figure} 
To reveal the macroscopic influence of cells on the droplet generation mechanism, we construct a phase diagram comparing the flow regimes of cell-laden and pure (cell-free) droplets. As shown in Fig. \ref{fig:8}, with the flow rate ratio $Q$ on the $x$-axis and the capillary number $Ca$ on the $y$-axis, the diagram shows the critical boundary transitioning from the squeezing-dominated DCJ regimes to the shear-dominated DC regimes.

A comparison of the fitted curves found that, at a fixed capillary number $Ca$, the transition boundary for cell-laden droplets (red dashed line) shifts notably towards lower flow rate ratios compared to that of pure droplets (blue dotted line). This implies that the presence of cells triggers the transition from DCJ to DC regime at a lower $Q$. For instance, at $Ca = 0.004$, the pure droplet system maintains the DCJ regime up to $Q \approx 0.9$, whereas the cell-laden system has already entered the DC regimes at $Q = 0.7$.
  
The physical origin of this shift lies in the “geometric blockage effect” induced by the cells. At the microchannel scale, cells act as solid cores, occupying a significant portion of the cross-sectional area. Following mass conservation, as the continuous phase flows through the confined gap between the cell and the channel wall, the effective hydraulic diameter of the channel decreases, thus increasing local flow velocity. This accelerated flow enhances the viscous shear stress exerted on the interface. Consequently, even at a lower bulk flow ratio, the intensified local shear might be sufficient to overcome the interfacial tension, leading to premature neck rupture and inducing an earlier transition in flow regimes.

Furthermore, we investigate the sensitivity of this phase boundary to cell biomechanics by varying the elastic modulus $C_{1}$ (300–1000 Pa). The results show that the mode transition points with varying cell stiffnesses are highly overlapping, collapsing onto a single curve (not shown here). This phenomenon emphasizes that the dominant mechanism is the geometric blockage effect rather than cell deformation. Although softer cells undergo greater deformation, their primary role during the critical pinch-off stage is to impose steric hindrance on the flow, thereby accelerating the local shear.

This insensitivity to cell stiffness has significant practical implications. It suggests that the boundary line possesses strong robustness, depending primarily on the blocking ratio $\Gamma$. Consequently, this phase diagram provides a universal predictive tool for microfluidic design, allowing parameter selection without the need for recalibration across different cell types or stiffnesses.

\subsection{\label{sec:level2}Effects of Cellular Heterogeneity}

To address the influence of cellular heterogeneity in microfluidic applications, this section investigates the mechanical interplay between cells and droplet dynamics. By independently regulating the cell blockage ratio $\Gamma$ and the elastic modulus $C_{1}$, we examine how variations in cell size and stiffness modulate the hydrodynamics of the encapsulation process and the resulting stress responses.
  
\noindent{\bf Cell blockage ratio $\Gamma$}

The cell blockage ratio $\Gamma$ serves as a fundamental dimensionless parameter characterizing the geometric confinement of cells within the microchannel. It quantifies the degree of spatial obstruction or the “blockage effect” imposed by the cells relative to the channel dimensions.

\begin{figure}
    \centering
    \captionsetup{justification=raggedright,singlelinecheck=false}
    \includegraphics[width=1\linewidth]{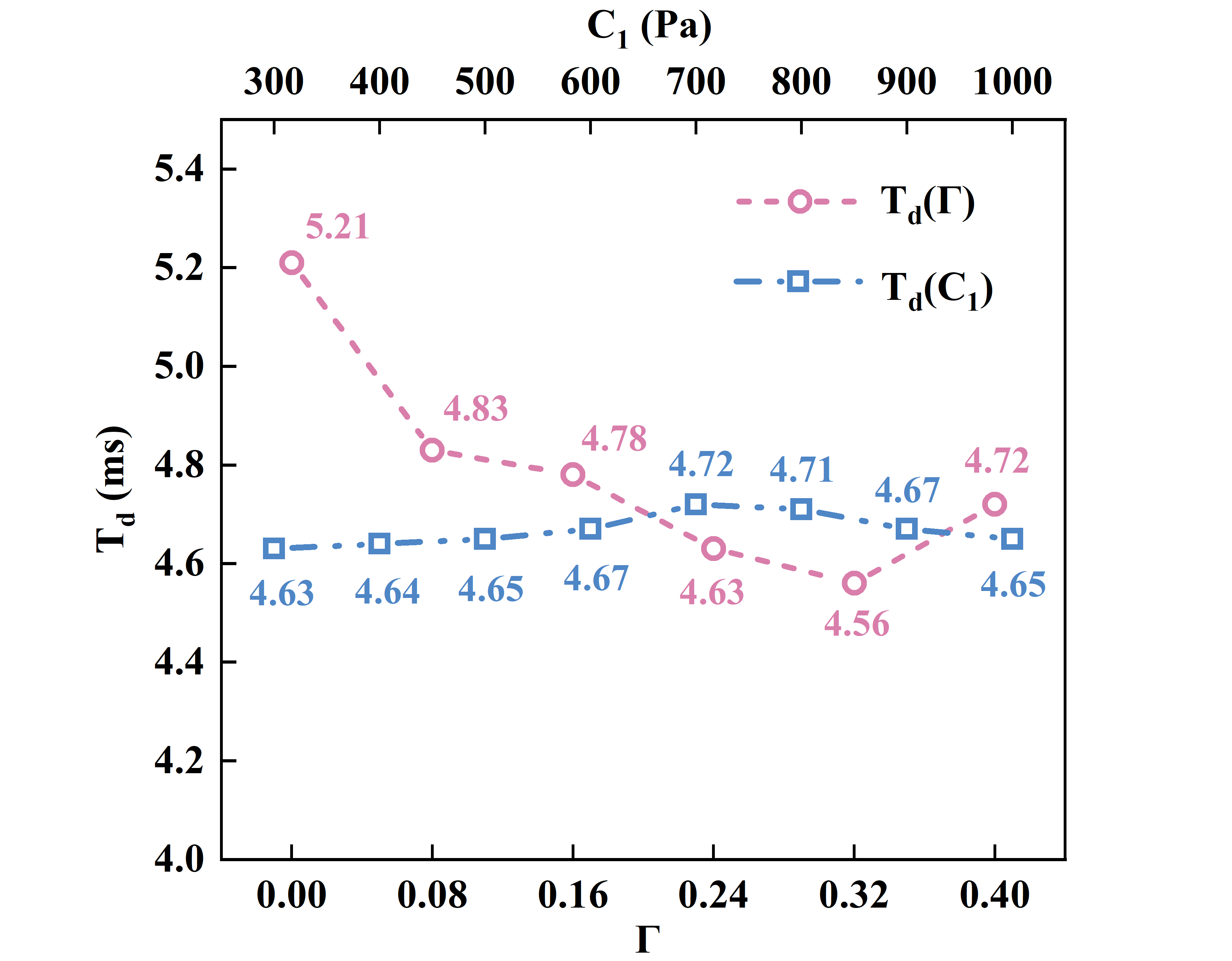}
    \caption{Variation of droplet generation period $T_d$ with cell blockage ratio and cell stiffness on droplet cycle. The flow parameters $Ca$ = 0.0018, $Re$ = 0.38, $Q$ = 0.5, $\Lambda$=1, $\lambda$=0.3, $L_{1}$ = 200 $\mu$m, $L_{2}$ = 600 $\mu$m, and $D_{c}$ = 84 $\mu$m. $C_{1}$ = 500 Pa is fixed for  $T_d\sim \Gamma$ curve, while $\Gamma$ = 0.40 is fixed for  $T_d\sim C_1$ curve.}
    \label{fig:9}
\end{figure}

Simulation results indicate a non-monotonic relationship between $\Gamma$ and the droplet generation dynamics (Fig. \ref{fig:9}). As $\Gamma$ increases from 0 (pure droplet case) to 0.32, the droplet generation period $T$ decreases significantly. However, at a higher blockage ratio of $\Gamma$ = 0.40, the period slightly rises.

\begin{figure*}[ht]
    \centering % 整个大组图在页面中居中
    \begin{subfigure}{0.32\textwidth}
        \centering
        \includegraphics[width=\textwidth]{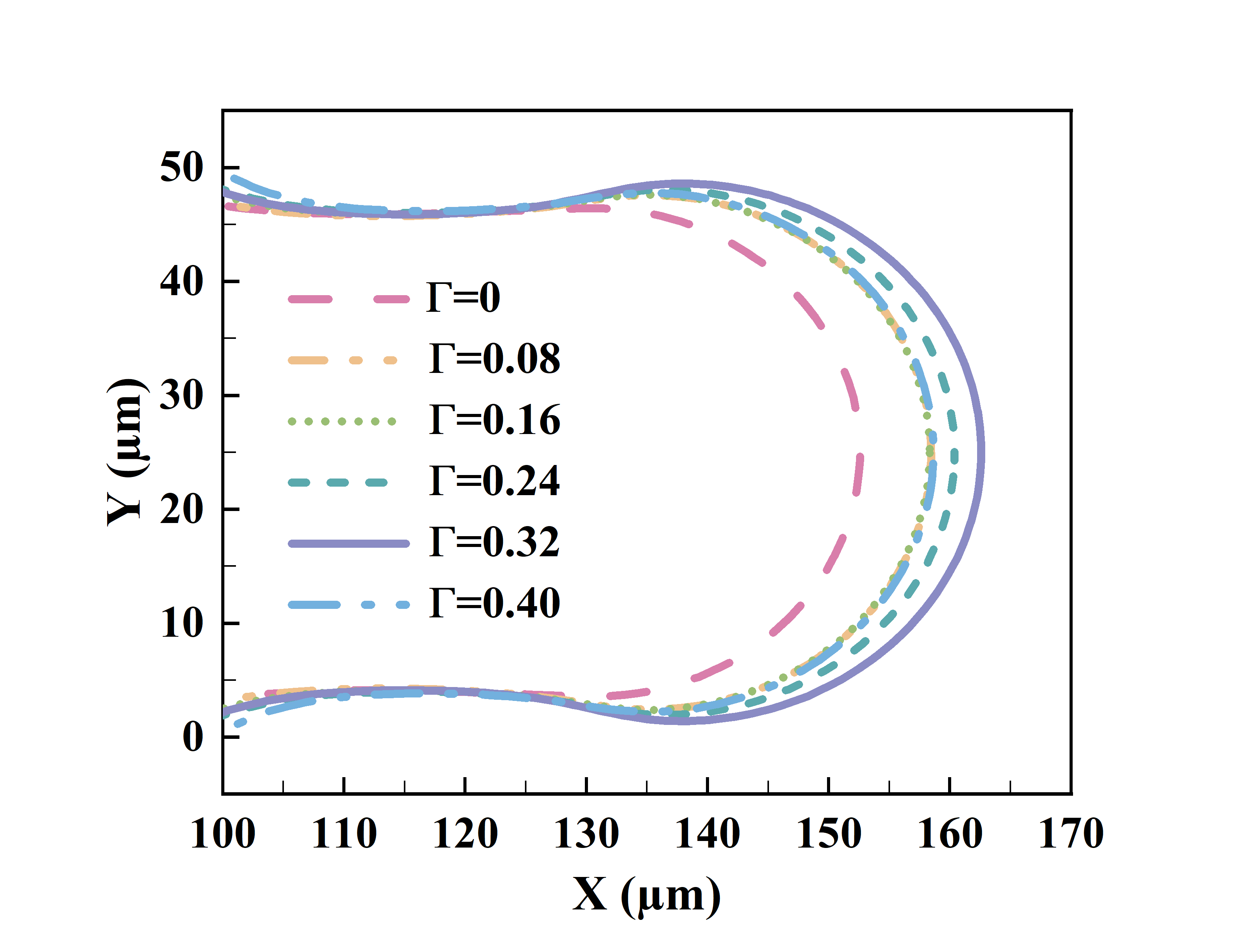}
        \caption{}
        \label{fig:10a}
    \end{subfigure}
    \hfill % 子图之间的空白
    \begin{subfigure}{0.32\textwidth}
        \centering
        \includegraphics[width=\textwidth]{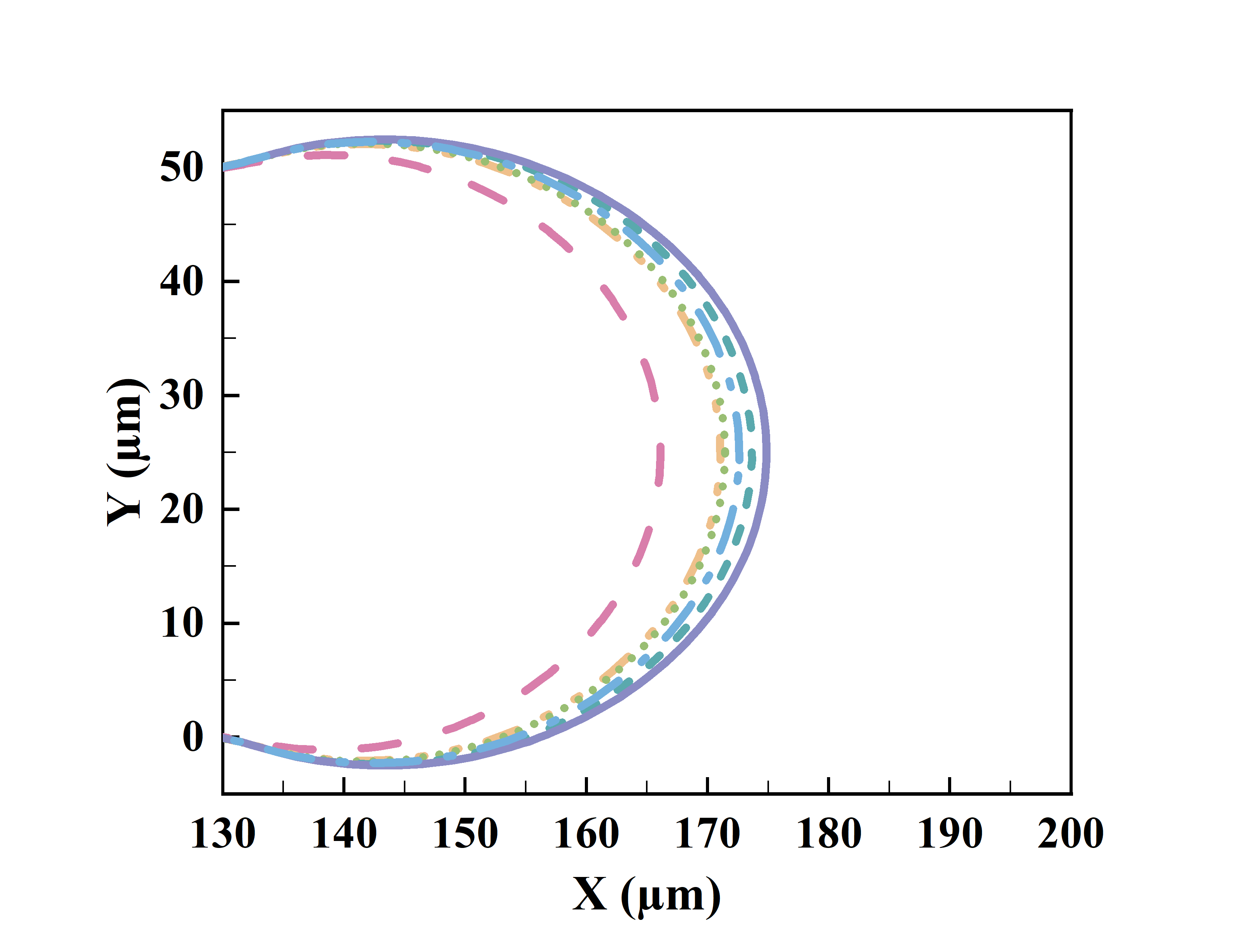}
        \caption{}
        \label{fig:10b}
    \end{subfigure}
    \hfill
    \begin{subfigure}{0.32\textwidth}
        \centering
        \includegraphics[width=\textwidth]{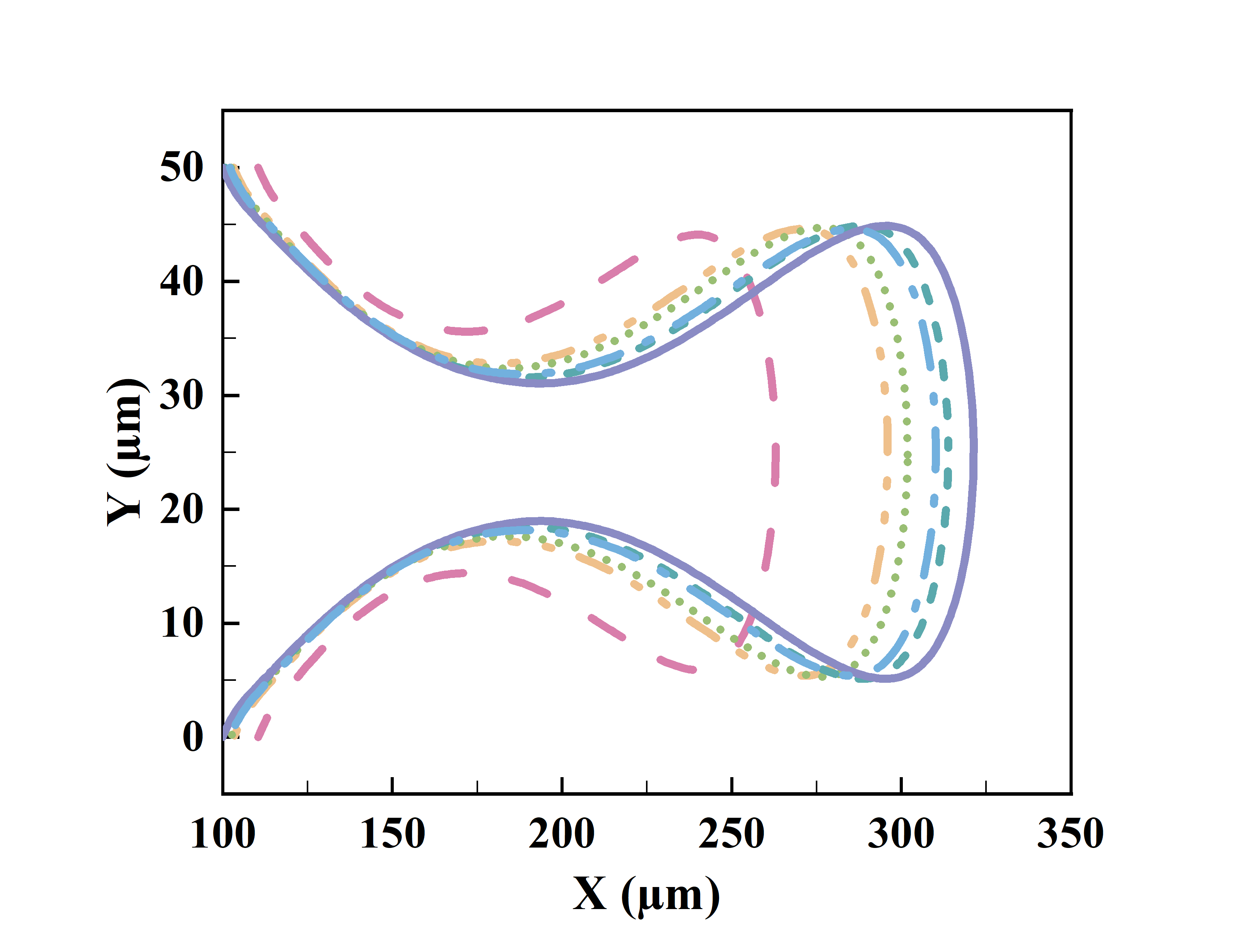}
        \caption{}
        \label{fig:10c}
    \end{subfigure}
    \captionsetup{justification=raggedright,singlelinecheck=false}
  % 整个大组图的标题
    \caption{The temporal evolution of the two-phase interface under different cell blockage ratios, where X and Y represent the axial and lateral positions, respectively. (a) $t$ = 1.5 ms (Retraction stage); (b) $t$ = 3 ms (Filling stage); (c) $t$ = 5.3 ms (Necking stage). Simulation parameters are $Ca$ = 0.0018, $Re$ = 0.38, $Q$ = 0.5, $\Lambda$=1, $\lambda$=0.3, channel geometry $L_{1}$ = 200 $\mu$m, $L_{2}$ = 600 $\mu$m, $D_{c}$ = 84 $\mu$m, and cell properties $C_{1}$ = 500 Pa, $C_{2}$ = 200 Pa.}
    \label{fig:10}
\end{figure*}

To visualize the impact of cell blockage on interface dynamics, we plot the spatiotemporal evolution of the two-phase interface (Fig. \ref{fig:10}). The interface for $\Gamma$ = 0.32 demonstrates the maximum axial displacement (frontmost position) during the droplet retraction, filling, and necking stages, indicating it has the highest interface propagation velocity. This corresponds to the most rapid transition from nucleation to critical detachment, yielding the shortest droplet generation time $T_d$. 
%The interface propagation speeds follow the order: $\Gamma$ = 0.32 > 0.24 > 0.40 > 0.16 > 0.08. 
Conversely, for the pure droplet case ($\Gamma=0$), the absence of a solid core prevents the geometric confinement and the local velocity acceleration effects, thus leading to the longest generation cycle.

These findings indicate that the non-monotonic trend of the droplet generation period $T_d$ is governed by a competitive mechanism between two opposing effects induced by cell blockage. On the one hand, the "shear enhancement effect" tends to shorten the generation period. As the cell occupies the channel, the reduced effective cross-sectional area accelerates the local flow velocity. This increases the viscous shear force exerted by the continuous phase, driving the droplet neck to thin and pinch off more rapidly. On the other hand, the "hydraulic resistance effect" tends to prolong the generation period. The narrowing gap between the cell and the channel wall inevitably increases the hydraulic resistance, which acts as a drag force that delays the advancement of the two-phase interface and the droplet filling process.

At $\Gamma = 0.32$, the system achieves an optimal trade-off where the shear-induced acceleration balances the resistance-induced deceleration. 
Conversely, at $\Gamma = 0.40$, the excessive resistance arising from the narrow gap overwhelms the shear acceleration effect, ultimately "braking" the interface and prolonging the period $T_d$.

\noindent{\bf Cell stiffness $C_{1}$}

It is found that, in the microfluidic cell encapsulation system, cell stiffness, a parameter of intrinsic mechanical properties, has a slight impact on the generation of the droplets. As shown in Figure \ref{fig:10}, the droplet generation period $T_d$ remains almost constant across varying cell stiffness parameters ($C_{1}$), ranging from 4.63 ms to 4.78 ms. The fluctuation amplitude is merely 0.09 ms, corresponding to a relative variation of approximately 1.9\%.

In contrast, hydrodynamic loads imposed on the cells are highly sensitive to cell stiffness heterogeneity. Under identical flow conditions, cells display differential mechanical responses, specifically in deformation and strain energy accumulation, due to their different constitutive properties. These variations critically impact structural integrity, ultimately determining post-encapsulation viability.

To quantify these effects, we vary the stiffnesses of the cells and investigate their influences on the cell's volumetric strain and strain energy density. The strain energy density function is described by the Mooney-Rivlin (M-R) model (Eq. (\ref{MR})). The volumetric strain $\theta$, a dimensionless quantity reflecting the degree of expansion or compression, is defined as the ratio of the change in volume to the initial volume:

\begin{equation}
\theta = \frac{V - V_0}{V_0},
\end{equation}
where $V$ represents the current volume of the cell, and $V_0$ represents the initial volume.
\begin{figure*}[ht]
    \centering % 整个大组图在页面中居中
    \begin{subfigure}{0.48\textwidth}
        \centering
        \includegraphics[width=\textwidth]{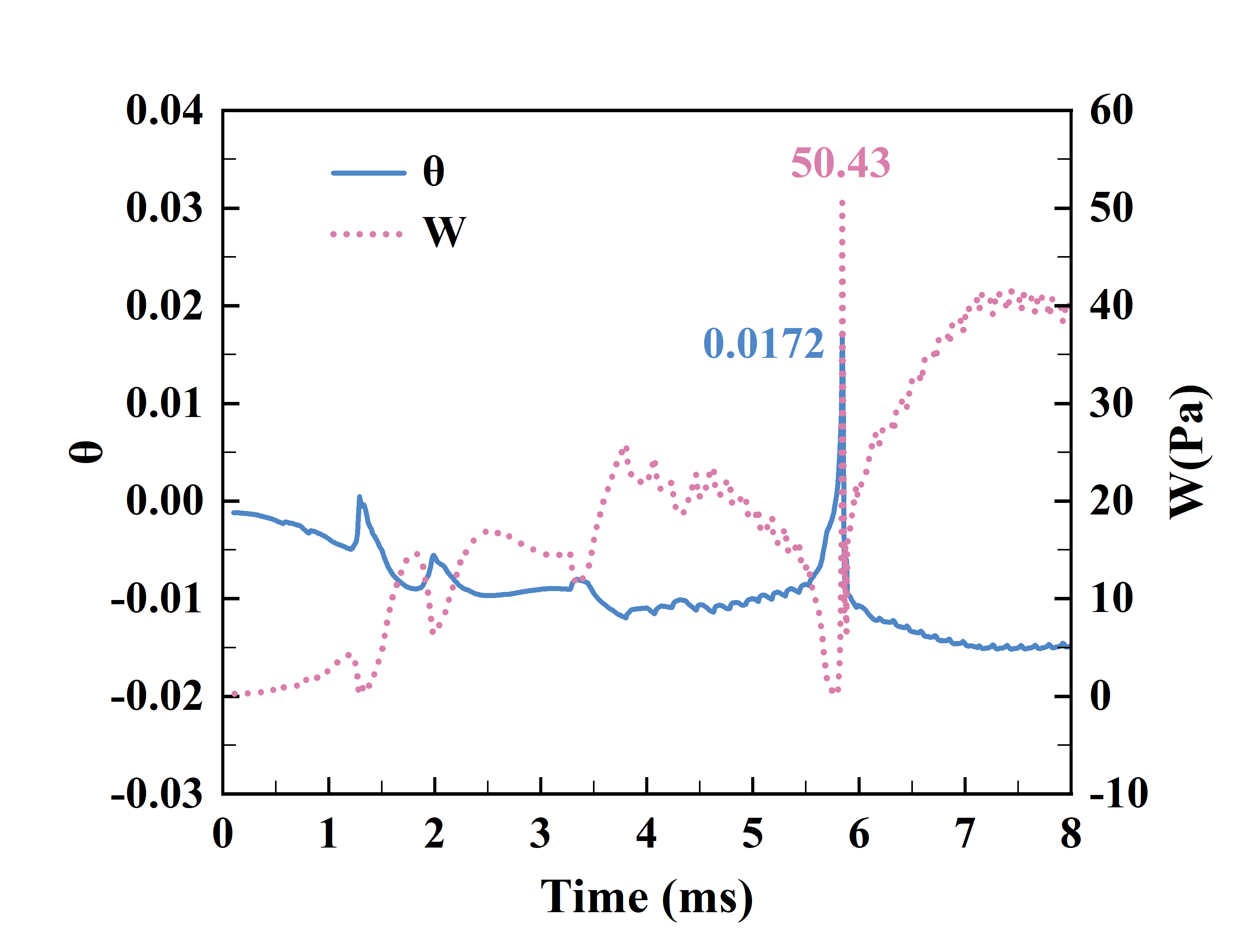}
        \caption{}
        \label{fig:11a}
    \end{subfigure}
    \hfill % 子图之间的空白
    \begin{subfigure}{0.48\textwidth}
        \centering
        \includegraphics[width=\textwidth]{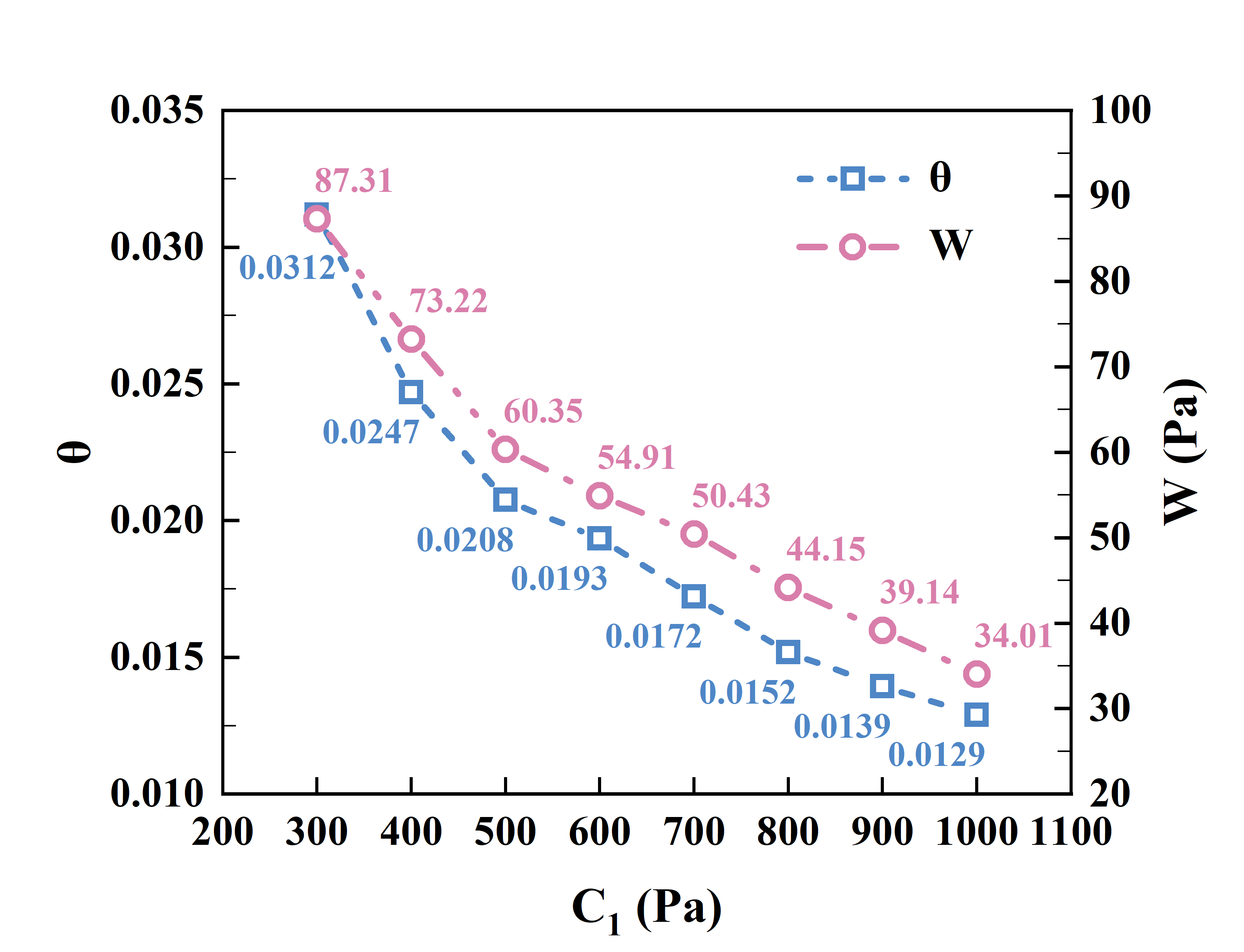}
        \caption{}
        \label{fig:11b}
    \end{subfigure}
  % 整个大组图的标题
   \captionsetup{justification=raggedright,singlelinecheck=false}
    \caption{Dynamic mechanical response of the cell. (a) Temporal evolution of volumetric strain $\theta$ (blue solid line) and strain energy $W$ (pink dotted line) at $C_1 = 500$ Pa. (b) Dependence of peak mechanical response ($t=5.8$ ms) on cell stiffness $C_1$. Other parameters are $Ca = 0.0018, Q = 0.5, \Gamma = 0.40, C_2 = 200$ Pa.}
    \label{fig:11}
\end{figure*}

Fig. \ref{fig:11}(a)illustrates the dynamic evolution of the cell's mechanical response. From $t = 0$ to 5.8 ms, both volumetric strain and strain energy demonstrate fluctuating characteristics, due to the dynamic coupling between hydrodynamic shear forces and the cell's intrinsic stiffness $C_{1}$. At $t = 5.8$ ms, the moment of droplet pinch-off, the instantaneous hydrodynamic impulse triggers a significant jump in the mechanical response, causing volumetric strain and strain energy to reach their peak values. At this critical instant, the mechanical loads imposed on the cell membrane and cytoskeleton approach their structural limits, representing a key window for assessing the mechanism of cell fragility. By examining these peak values across different stiffness parameters, we can identify cells that are fragile to rupture due to mechanical overload. As shown in Fig. \ref{fig:11}(b), an increase in stiffness $C_{1}$ enhances the cell's resistance to deformation. Under identical fluid loads, stiffer cells display reduced maximum volumetric deformation and lower peak strain energy. This negative correlation provides a quantitative mechanical basis for explaining the susceptibility of softer cells to mechanical damage.

\section{Conclusion}

In the present study, a numerical framework that couples the multiphase flow model with the Arbitrary Lagrangian-Eulerian (ALE) fluid-structure interaction method is employed to conduct a comprehensive investigation into the dynamic encapsulation of single hyperelastic cells in microfluidic droplets. Unlike traditional experimental approaches, this simulation provides efficient access to the transient variations in velocity and mechanical stress fields acting on cells, offering deep insights into the fluid-solid coupling mechanisms. 

We identify three encapsulation modes governed by the initial cell position, namely "Normal Encapsulation" (Zone III), "Lagging Pinching" (Zone II), and "Premature Escape" (Zone IV). Successful encapsulation requires a precise spatiotemporal match between cell transport and droplet pinch-off. Importantly, encapsulation failures not only result in trapping defects but also subject the cells to excessive mechanical stress, significantly increasing the risk of biological damage. We derive a unified scaling law that quantitatively defines the safe operational boundaries.

The physical presence of cells will change the macroscopic flow regimes of droplet generation. Due to the geometric blockage effect, the transition boundary from the squeezing-dominated (DCJ) to the shear-dominated (DC) mode shifts toward the lower flow rate ratio $Q$ region. This implies that cell-laden droplets enter the shear-dominated regime earlier than pure droplets under identical flow conditions.

The droplet generation period demonstrates a non-monotonic dependence on the cell blockage ratio $\Gamma$. Simulation results reveal that due to the competitive mechanism between shear enhancement due to reduced cross-sectional area and hydraulic resistance due to narrower gaps, an optimal balance is achieved at $\Gamma \approx 0.32$, where the interface propagation is fastest. Beyond this critical range (e.g., $\Gamma = 0.40$), excessive hydraulic resistance overwhelms the shear benefit, prolonging the generation period.

It is found that macroscopic droplet generation is robust to cell stiffness $C_1$. However, the mechanical response of the cell itself to the flow field is highly sensitive. Specifically, results indicate that the capillary necking (pinch-off) stage is a singularity point for mechanical loading, in which cells experience peak hydrodynamic forces. Softer cells accumulate higher strain energy and undergo larger deformation, making them more vulnerable to mechanical damage. Crucially, our simulation overcomes the inherent difficulty of experimental techniques in measuring these real-time, in-situ dynamic stresses, providing a quantitative basis for assessing cell viability risks.

By resolving the multiscale problem involving interface dynamics and solid deformation, this work explores the transient stress singularities inaccessible to current experimental techniques. The established scaling laws and geometric blockage mechanism enrich the fundamental theory of particle-laden multiphase flows. This study also provides a quantitative, mechanics-based framework for optimizing the design of microfluidic systems and achieving high-fidelity, damage-free encapsulation. 

\section*{Acknowledgement}

This research is supported by the National Natural Science Foundation of China (No. 12332016) and the Science and Technology Commission of Shanghai Municipality (Grant No. 24TS1412500). 
%Beijing Beilong Supercloud Computing Co., Ltd. provides support in computational resources.

\section*{references}

\nocite{*}
\bibliography{aipsamp}% Produces the bibliography via BibTeX.

\end{document}